\documentclass[sigconf,screen]{acmart}

\usepackage{subcaption}
\usepackage[table]{xcolor}

\usepackage{academicons}
\usepackage{xcolor}

\AtBeginDocument{%
  \providecommand\BibTeX{{%
    \normalfont B\kern-0.5em{\scshape i\kern-0.25em b}\kern-0.8em\TeX}}}

\setcopyright{acmlicensed}
\acmDOI{123456789}
\acmYear{2026}
\copyrightyear{2026}
\acmSubmissionID{not-yet-known}
\acmISBN{979-8-4007-0658-5/24/07}
\acmConference[FSE Companion '26]{Companion Proceedings of the 34th ACM International Conference on the Foundations of Software Engineering}{June}{Montreal}
\acmBooktitle{Companion Proceedings of the 34th ACM International Conference on the Foundations of Software Engineering (FSE Companion '26), June, 2026, Montreal}
\usepackage{tabularx}
\begin{document}

\newcommand{\orcidicon}[1]{}

\title{Just-in-Time Catching Test Generation at Meta}
\author{
Matthew Becker\orcidicon{0009-0005-9860-3583},
Yifei Chen\orcidicon{0009-0009-1059-933X},
Nicholas Cochran\orcidicon{0009-0001-5212-3817},
Pouyan Ghasemi\orcidicon{0009-0008-1685-8067},
Abhishek Gulati\orcidicon{0009-0008-5347-5321},
Mark Harman\orcidicon{0000-0002-5864-4488}\textsuperscript{*},
Zachary Haluza\orcidicon{0009-0008-0272-7808},
Mehrdad Honarkhah\orcidicon{0009-0006-4172-2124},
Herve Robert\orcidicon{0009-0005-9727-4562},
Jiacheng Liu\orcidicon{0000-0002-1449-5038},
Weini Liu\orcidicon{0009-0007-0604-6842},
Sreeja Thummala\orcidicon{0009-0008-4848-5429},
Xiaoning Yang\orcidicon{0009-0002-5929-5627},
Rui Xin\orcidicon{0009-0003-9563-1242},
Sophie Zeng\orcidicon{0009-0003-5012-6699}
}
\affiliation{Meta Platforms, London UK, Seattle and California \country{USA}}
\thanks{\textsuperscript{*}The corresponding author is Mark Harman. Author order is alphabetical.}

\renewcommand{\shortauthors}{Becker et al.}

\newcommand{\DPRSThreshold} {20\%}
\newcommand{\NumberOfReachouts} {41} %%% Mark computed this by searching for Workchat threads which involved at least five participants and included Mark, Sreeja, Sophie and Mehrdad (and involve the word expected). He then checked each one to make sure it was one of these reach out workflows. This includes the DRS and DPRS reached outs. We could improve our figures by focusing solely on DPS, but I think that's somewhat cheating and I'd rather quote the worst figure for overall true positive to be completely fair and honest. 
\newcommand{\NumberOfDiffsInTPValidation} {3,488}
\newcommand{\SpreadsheetWeakCatchesForLLMJudge} {600}
\newcommand{\SpreadsheetWeakCatchesForLLMJudgeManuallyAnalysed} {600}  % confirm final number after cleaning up spreadsheet
\newcommand{\OverallEngineerPerceivedTPRate}{19.5\%}
\newcommand{\highestRankCorrlection}{0.535}

\begin{abstract}
We report on Just-in-Time catching test generation at Meta, designed to prevent bugs in large scale backend systems of hundreds of millions of line of code. 
Unlike traditional hardening tests, which pass at generation time, catching tests are meant to fail, surfacing bugs before code lands. 
The primary challenge is to reduce development drag from false positive test failures. 
Analyzing 22,126 generated tests, we show code-change-aware methods improve candidate catch generation 4x over hardening tests and 20x over coincidentally failing tests. 
To address false positives, we use rule-based and LLM-based assessors.
These assessors reduce human review load by 70\%. 
Inferential statistical analysis showed that 
human-accepted code changes are assessed to have significantly more false positives, while human-rejected changes have significantly more true positives. 
We reported \NumberOfReachouts~candidate catches to engineers; 
8 were confirmed to be true positives, 
4 of which would have led to serious failures had they remained uncaught. 
Overall, our results show that Just-in-Time catching is 
scalable, 
industrially applicable, 
and 
that it prevents serious failures from reaching production.
\end{abstract}

% %%
% %% The code below is generated by the tool at http://dl.acm.org/ccs.cfm.
% %% Please copy and paste the code instead of the example below.
% %%
\begin{CCSXML}
<ccs2012>
<concept>
<concept_id>10011007.10011074.10011099.10011102.10011103</concept_id>
<concept_desc>Software and its engineering~Software testing and debugging</concept_desc>
<concept_significance>500</concept_significance>
</concept>
</ccs2012>
\end{CCSXML}

\ccsdesc[500]{Software and its engineering~Software testing and debugging}
\keywords{Unit Tests, Automated Testing, LLMs, Test Oracles. }

\maketitle

\section{Introduction}

Most work on unit test generation 
\cite{mhetal:fse24-llm,anand2012automated,fraser:evosuite,mcminn:past-present-and-future,mh:icst15-keynote,ryan2024code,chen2024chatunitest,schafer2023empirical,liu2024llm} is concerned with `hardening'  tests.
That is, tests that pass at generation time, and land 
into the code base to `harden' against future regressions.
By contrast, `catching' tests {\em fail} at test generation time 
by design, with the aim of catching bugs in proposed changes \cite{mhpohss:harden}.
A catching test cannot land together with the changes that it tests, since it fails on those changes.
Therefore, the catching test must either be discarded (because it is deemed unhelpful), 
or its signal must be addressed (to fix the changes and make the test pass).

We target \emph{regression} catching tests that fail on the proposed changes (a code revision) but pass on the parent revision.
For catching tests we cannot rely on the \emph{implicit oracle}~\cite{ebetal:oracle}.
Our auto-generated tests must detect regressions according to the \emph{general oracle}—the true, expected behavior, which is often ill-defined and only partly knowable~\cite{ebetal:oracle,weyuker:untestable}. 
The implicit oracle merely captures errors, such as crashes, that are known to be true positive failures regardless of the specification~\cite{ebetal:oracle,goffi:automatic}.
By contrast, the general oracle is the (usually vague and informal, even unstated) specification of correct behavior.
This inherent vagueness makes it hard (yet no less important) to distinguish true from false positive test failures. 

A weak catch is simply one that fails on the current revision, 
whereas a {\em strong} catch is one that also {\em should} fail (a true positive failure) according to this general oracle.
Its failure reveals a bug in the code under test.
A {\em strictly} weak catch is a false positive failure that reveals a bug in the test case, including oracle misalignment.
The  challenge is to determine whether a weak catch is strong.
The problem of automatically generating such strong catching tests Just-in-Time (JIT) as revisions are submitted, is called the `Catching JiTTest Challenge', introduced in the FSE 2025 keynote paper ~\cite{mhpohss:harden}.

At Meta, a revision submitted to the Continuous Integration (CI) system  is called a `diff' (short for differential).
Our catch generation workflows are deployed at diff submission time.
We focus on the potential for a diff to introduce a {\em severe} regression.
It is important to focus on severe regressions, because catching test generation is inherently computationally expensive; many tests must be generated to increase the chances of finding a strong catch.
We prioritize catching severe bugs using a targeting mechanism similar to Diff Risk Score (DRS) \cite{abreu:DRS}, trained on previous code changes and running overnight (west coast USA) using spare capacity.
The workflow is triggered for all the diffs submitted for the previous day that are flagged as high risk by the targetter.
We seek to answer the following question for automated software test generation:

\begin{quote}
\it
    How can we automatically generate tests that find the most severe bugs in proposed changes, before the bugs cause severe failures and with high enough true positive rate that the test signal does not impede engineer velocity?
\end{quote}

We deployed two catch generation workflows and three true/false positive assessors. 
We used them to test Meta's systems of hundreds of millions of lines of code used by over 3.5 billion people worldwide every day. 
Our primary findings are as follows:-

\noindent
{\bf Effectiveness of diff-awareness}:
    Based on a study of  
    22,126
    tests generated, we found that diff-aware  catch generation workflows produce approximately 4 times as many weak catches as hardening workflows. 
    They generate 20 times as many as a baseline that simply attempts to find tests that fail on the diff, and coincidentally pass on its parent revision.

\noindent
{\bf Scaling Human-in-the-loop}:    
    Based on a manual study of \SpreadsheetWeakCatchesForLLMJudge~ diffs with weak catches,  we found that the LLM-as-judge ensemble helped reduce human-in-the loop review load by approximately 70\%, allowing us to scale evaluation capacity to approximately 4 times that of manual review alone.

\noindent
{\bf Consistent with human labelling}:
    Based on a study of diffs reviewed by humans, and thereby labelled as `Good' (accepted to land into the code base) or `Bad' (not accepted to land), we found  statistically significantly more false positive assessments for tests that weakly catch diffs labelled `Good' and significantly more true positive assessments for those labelled `Bad'. 
    We also found  only modest inter-rater agreement and rank correlation between LLM-based and rule-based assessors.
    Taken together, these findings suggest that the assessors provide complementary discriminating signals.

\noindent
{\bf Severe failures averted}:    
    Finally, and most importantly, from \NumberOfReachouts~reach outs to engineers based on  failing test signals from catch, 8 were immediately confirmed to be strong catches, yielding an engineer-experienced true positive rate of \OverallEngineerPerceivedTPRate. 
    Of the 8 strong catches, 4 would have led to serious failures in production. 
    These 4 serious failures were thus averted by Catching JiTTests.
    We also found that dismissing false positives took only a few minutes and therefore had negligible drag on developer velocity.

\section{Baselines for Just in Time Catching}
A prerequisite for any approach to generate strong catches on a diff is the ability to generate weak catches.
In order to assess the degree to which existing technologies can generate weak catches, 
we used three `baseline' approaches, described below.

\subsection{Coincidental regression catches found during ACH hardening deployment}
We have already deployed mutation-guided LLM-based hardening test generation \cite{foster:mutation} into Meta's continuous integration system, Phabricator~\cite{feitelson:deployment}.
As a byproduct of this hardening test generation deployment, 
we also produce tests that fail on the diff under test.
According to the mutation-guided workflow~\cite{foster:mutation} these are discarded since they cannot land, by construction.
However, where such tests also happen to pass on the parent of the tested diff, these tests denote a weak catch.
We call these catches `coincidental catches', because the goal of the test generation was not to produce a weak catch, but to produce a strong hardening test.

\subsection{Hardening test generation workflows as catching workflows}
To provide additional baselines on the chance of generating a weak catch using existing approaches, we also collected results from the application of TestGen-LLM~\cite{mhetal:fse24-llm}, and ACH \cite{foster:mutation} to the parent of the tested diff.
TestGen-LLM~\cite{mhetal:fse24-llm} is a traditional LLM-based test generation system that elevates coverage, stimulating the LLM to generate test cases to target code that is currently uncovered by any existing test case. 
The system is based on line coverage and has been deployed (and regularly updated and evolved) at Meta for two years now.  
For establishment of a hardening-based baseline, we use the original system as reported in the literature~\cite{mhetal:fse24-llm}.
ACH is a mutation-based test generation system \cite{foster:mutation}
that generates mutants and test cases that kill them.

These two approaches are both deployed at Meta to generate hardening test cases. 
These are merely `baseline' approaches, because they use existing technologies `out of the box'.
When such a test case, generated for the parent, fails on the child, it denotes a `Weak Catch' as well as a `Harden' according to the `Harden and Catch` Framework~\cite{mhpohss:harden}.

For the TestGen-LLM workflow, we also experimented by removing the requirement to construct tests that achieve additional coverage. 
That is, we simply require that the workflow should generate a test that builds and passes on the parent revision. 
This gives the hardening workflow the greatest chance of generating a valid catching test,
since we know that approximately half the tests that find simulated bugs do {\em not} improve coverage~\cite{foster:mutation}.

Neither of these hardening workflows are diff–aware. 
That is, the diff is the current revision, but each of the two workflows is applied, not to the {\em current} revision, but to its parent
(the base revision on which the diff is based).
Therefore, neither workflow is supplied with any information about the diff itself. 
In their production deployment as `hardeners', they cannot `know' about the diff, 
since they have to cater for {\em any}  future (unknown) changes.

\section{Diff-Aware Workflows}
\label{sec:workflows}

\begin{figure*}

%\centerline{\includegraphics[width=1.25\linewidth, trim={1cm 0 0 5cm}, clip]{figures/architecture.pdf}}
\centerline{\includegraphics[width=1\linewidth]{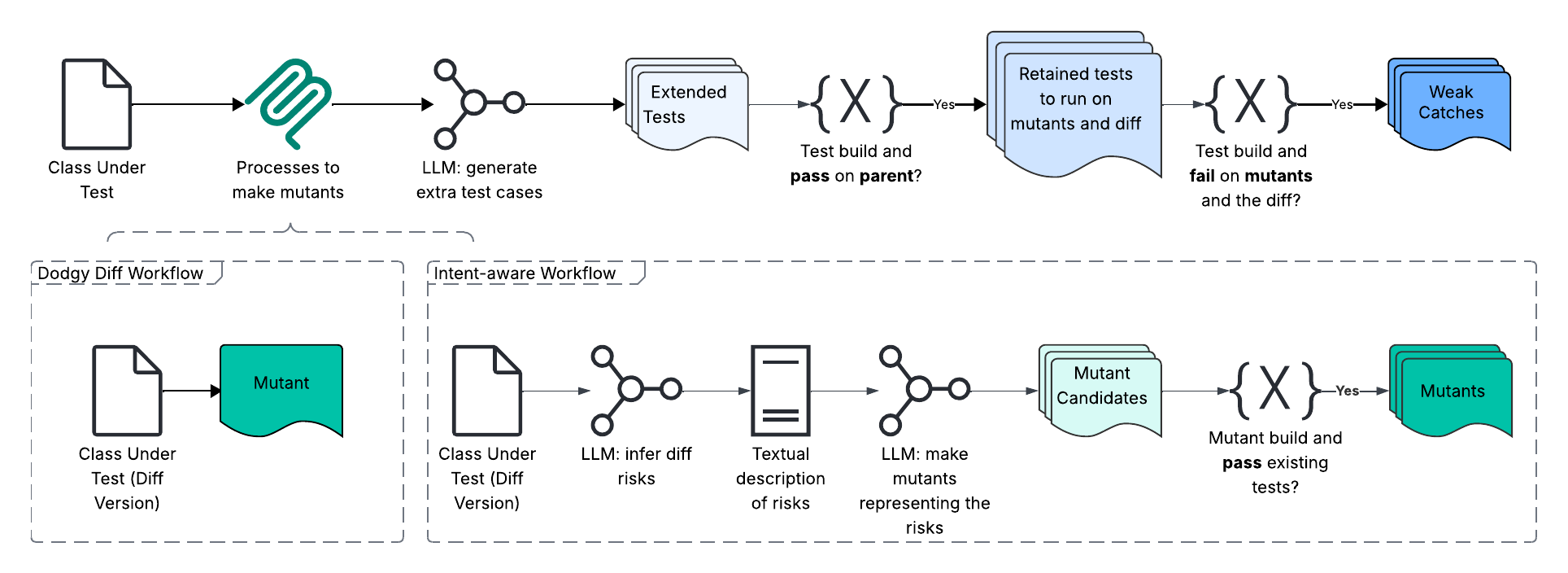}}
\vspace{0.0cm}
\caption{Architecture of `Dodgy diff' and Intent-Aware  Workflows for generating Just-in-Time Catches.}
\label{fig:workflows}
\end{figure*}

Unlike hardening workflows, a catch generation workflow can always choose to exploit information about the diff to help it find ways 
to make the test fail on that diff.
A purpose-built catch workflow is  thus `diff-aware'.
We have experimented with two approaches to generating weak catches, based on this diff-awareness (both depicted in Figure~\ref{fig:workflows}). 
The `dodgy diff' workflow is aware of the diff, but makes no attempt to infer anything about its intention.
Rather, it treats the diff as if it were a mutant of its parent, implicitly asserting that the diff is `buggy' (hence the `dodgy diff' moniker).

We know, from previous studies, that the vast majority of code changes do not induce failures (a corollary of the competent programmer hypothesis~\cite{yjmh:analysis}).
Nevertheless, it makes sense to have a workflow that maximizes the number of weak catches and thereby maximizes the pool of candidates from which to find a strong catch.
The dodgy diff approach relies upon the existence of post-processing to weed out false positive failures, a topic to which we will return Section~\ref{sec:judges}.

We also want a workflow that favors the generation of tests that are more {\em likely} 
to reveal genuine bugs rather than to simply elevate the chances of generating a weak (but not necessarily strong) catch. 
We therefore implemented a `diff-intent aware' workflow. 
This workflow attempts to approximate the {\em intention} of the diff.
We first describe the intent-aware workflow, and then the dodgy diff workflow, 
which is a simply a projection of the intent-aware workflow onto a simple subset.
Both approaches are inspired by mutation testing \cite{yjmh:analysis}.

\subsection{Intent aware workflow}
The intent-aware workflow is depicted in Figure~\ref{fig:workflows} (lower right-hand side).
The workflow first stimulates the language model to describe the ways in which an attempt to implement the intent could produce errors. 
We call this the `risks' of the diff. 
The workflow thereby becomes `risk aware' (based on its intent awareness).

From the description of these risks, we construct mutants. 
Each mutant implements a code modification, based on the parent of the tested diff. 
Each mutant captures a bug that {\em could} be introduced as the result of attempting to implement the diff intent.
Of course, we do not expect that we will necessarily catch a bug in the tested diff, 
because it may not precisely mirror  one of the inferred risks, even if it is buggy. 

Nevertheless, from the considerable empirical evidence collected (over multiple decades) on the mutation `coupling hypothesis' \cite{yjmh:analysis}, 
it is reasonable to suppose that real faults may be `coupled' to these mutants.
Therefore, capturing the inferred risks as mutants is a natural first step towards generating tests that check for each risk.

For each of the mutants, we generate a test following the Mutation-Guided LLM-based Test Generation workflow on the parent. 
That is, we seek to generate tests that will  pass on the parent but  would fail on the mutated parent version. 
These tests are then tested on the diff itself, and those that fail on the diff are harvested as weak catches.

\subsection{Intent-aware baseline for intent computation}
Computing the intention behind a code change is inherently problematic. 
We are essentially tackling, head on, the Oracle problem in software testing \cite{ebetal:oracle}. 
Before the advent of LLM-based Software Engineering~\cite{mhetal:LLM-survey}, 
it would have been presumed a foolhardy endeavor: how can we know what the intention of a code change is, when all we have is the code?

However, as noted previously \cite{mhpohss:harden}, there is often extra contextual information with a code change, such as the title, summary, emails, and discussion between engineers, task descriptions, and other documentation available.  
This contextual information can be used to approximate the intention underlying a code change.

We wanted to provide a baseline against which to measure such richer sources of information, and to understand the degree to which simple LLM code summarization could be used to approximate the intention of a diff.
With this in mind, the first version of the intent-aware workflow we deployed simply used the code itself, 
and the diff title and summary, and no other contextual information in order to infer the diff intent.

\subsection{Dodgy diff: intent-unaware workflow}
\label{sec:dodgy-diff}
The `dodgy diff' workflow is depicted in Figure~\ref{fig:workflows}(lower left-hand side).
As can be seen from Figure~\ref{fig:workflows}, the dodgy diff workflow is simply a projection of 
the intent-aware workflow. 
We simply treat the diff as if it were already a mutant (a buggy version of the parent).
We then use a Mutation-Guided LLM-based Test Generator \cite{foster:mutation} to generate tests that attempt to kill the mutant (that is, to distinguish its behavior from that of its parent).

\section{Results of applying the workflows}
Table~\ref{tab:numbers-of-weak-catch-results} presents the results for the two diff-aware workflows, the two hardening baselines, and the coincidental catch baseline.
The two diff-aware work flows are part of our production deployment since September 2025 and so results were simply taken from production for these.
The non-diff-aware workflows are not part of our production deployment and were run solely for the purpose of collecting the results for this paper.
We ran the baseline diff-unaware workflows for a period of 10 days between 14th and 24th November 2025.  
During the 10-day period of assessment, the workflows were applied to the same set of diffs\footnote
{
except for coincidental catches which were taken from production over a longer period, and were initially included as part of the initial production deployment, until 11th November 2025, when the coincidental catch workflow was discontinued.
} as the production workflows.
The total number of tests generated for the four workflows (two diff-aware and two diff-unaware) was 9,125.

\begin{table*}
    \caption{Results from five different approaches to generating weak catches as diffs are submitted}
    \label{tab:numbers-of-weak-catch-results}
    \centering
    \begin{tabular}{||l|r|r|r||r|r|r||}
        \hline
        \textbf{}         & \textbf{Total}     & \textbf{Total}   &  \textbf{\%age} & \textbf{Total }  & \textbf{Total diffs } &  \textbf{\%age}\\ 
        \textbf{}         & \textbf{Tests}     & \textbf{Weak }   &  \textbf{Weak}   & \textbf{Diffs With}         & \textbf{with Weak }   &  \textbf{Diffs}\\ 
        \textbf{Approach} & \textbf{generated} & \textbf{Catches} &  \textbf{Catch}      & \textbf{Tests}        & \textbf{Catches}      &  \textbf{Caught}\\ 
        \hline
        \hline     
                \multicolumn{7}{c}{ } \\  \hline
        \multicolumn{7}{||c||}{Weak {\bf catches found by coincidence} while attempting to generate hardening tests for the diff's parent} \\  \hline
        Coincidental catch                          &  13,001  &    29  & 0.2\%  & N/A   &    N/A  & N/A     \\  \hline
        \multicolumn{7}{c}{ } \\  \hline

        \multicolumn{7}{||c||}{hardening approaches that are {\bf  not aware of the diff under test}} \\  \hline
        %make\_test\_extensions\_on\_parent\_then\_retest\_on\_diff    
        TestGen-LLM~\cite{mhetal:fse24-llm}   -- need {\em not} add coverage     &  4,499  &    89  & 2.0\%  & 1,185  &  43  & 3.6\%   \\  \hline
        TestGen-LLM~\cite{mhetal:fse24-llm}   -- {\em must also} add coverage\footnote{These numbers are estimated based on subsequent competition of proportions of passing and failing tests that add coverage. 16\% of failing tests were found to add coverage, whereas 32\% of a same-sized random sample of passing tests were found to add coverage. 
        4,410  passing (1,440 estimated to add coverage) plus 89 failing (14 add coverage) so 1,454 add coverage overall. 
        }    
        &  1,440  &    14  & 0.9\%  & N/A  &  N/A  & N/A   \\  \hline
        %make\_and\_kill\_mutants\_on\_parent\_then\_retest\_on\_diff        
        ACH Mutation-Guided~\cite{foster:mutation}  &    754  &      6  & 0.8\%  &   322  &   5  & 1.6\%   \\  \hline
        \hline
        \hline

        \multicolumn{7}{c}{ } \\  \hline

        \multicolumn{7}{||c||}{Just-in-Time Catching  approaches that {\bf are  aware of the diff under test}} \\  \hline
        %dodgy diff workflow                 
        Dodgy diff                                  &  1,621  &    41  & 2.5\%  &   654 &    26  & 4.0\%   \\  \hline
        %intent-aware workflow      
        Use computed diff intention and risks       &    811  &    52  & 6.4\%  &   617 &     49  & 7.9\%   \\  \hline
        \hline
        Totals                                      &  22,126 &     231  & 1\%    & N/A   &    N/A  & N/A     \\ \hline 
        \hline
    \end{tabular}
\end{table*}

\subsection{Results for co-incidental catches}
The first row of results concerns coincidental catches. 
We report  the proportion of these `fail-on-diff' tests that also, coincidentally, pass on the parent of the diff under test, thereby constituting a weak catch. 
This proportion 
 is 0.2\%; noticeably lower than the other two baselines reported in the table.
The low baseline result for coincidental  catches 
highlights the challenges of finding them.
The noticeably lower results for the number of weak catches led us to abandon any further  attempt to harvest `co-incidental catching' as a practical source of weak  catches.

\subsection{Results for hardening baseline}
The TestGen-LLM approach generates 10 times as many catches as the coincidental catch approach.
However, it should be noted that this approach simply attempts to generate a test that will build and pass on the parent of the tested diff.
It makes no attempt to give any assurance that these tests add value.
If we focus more specifically on those that add coverage, the increase in the number of weak catches generated becomes 4x that for a coincidental catch, rather than 10x.
For the mutation guided approach, we observe a similar increase in the number of weak catches generated.

From these numbers, we can conclude that, if we generate a hardening test,
then
the probability that it proves to be a weak catch on the tested diff 
is approximately four times greater than the probability that a test that happens to fail on the diff will prove to be a weak catch.
These results reflect the relative chance of making a test that passes, 
compared to the chance of generating, for arbitrary code,  a test that fails.

% interesting, but can be deleted if we need space ...
The results for the hardening workflows also implicitly assess the probability of a hardening test failing on some possible future change.
Essentially, we have picked an arbitrary change, generated a hardening test on the fly, based on the parent of that change, 
and then assessed the chances that the hardening test would fail for the change.
From Table~\ref{tab:numbers-of-weak-catch-results} we can see that the probability is approximately 0.01 (0.008 for mutation-guided generation and 0.009 for coverage-guided generation).
As a rough overall expectation estimate, we therefore would expect hardening tests to fail on approximately 1\% of the diffs to which they are applied.

\subsection{Results for diff-aware weak catch generation}
We expect that if we implement {\em any} diff-aware catch generator, it should perform better than the hardening baselines,
because it generates tests for a specific (rather than an arbitrary future) code change.
Indeed, as the results from Table~\ref{tab:numbers-of-weak-catch-results} show, 
both diff-aware workflows generate far more weak catches per test generated, although fewer overall.

More importantly, the diff-aware approaches are far more capable of finding a  weak catch {\em per diff}.
That is, when we generate tests without any guidance (neither coverage not mutation) we get catches for 3.6\% of diffs, 
and with mutation-guidance this is 1.6\%.
By contrast, using the dodgy diff approach we generate weak catches for 4.0\% of diffs, rising to 7.9\% of diffs using the intent-aware workflow.
We conclude that, by taking advantage of information about the diff and an approximation of its intention, 
we can more than double the number of code changes for which we generate a weak catch.

Looking at the differences between the two diff-aware  catch approaches, we also see notable differences in the number of tests that are generated.
Although the intent-aware approach generates fewer tests than the `dodgy diff' approach,
the tests that it generates are far more likely to be weak catches, both per test and overall. 
More importantly, 
the intent-aware workflow is
able to produce at least one weak catch for almost double the number of diffs.
This provides  initial evidence that making the workflow intent aware, and focusing on likely risks, can significantly increase the chance of generating a weak catch.

We do not claim that our two diff-aware workflows are the best obtainable. 
These findings  merely provide a foundation for further research and development.
Much more research is needed to imbue diff-aware workflows with richer context allowing them to generate more catching tests, 
and more weak catching tests that are likely to also prove to be strong.
We present these results to provide a baseline for future research investigations on catching tests.

\section{Initial Evidence that  significant real world  bugs will be prevented by targeted catch} 
\label{sec:examples-of-bugs}
The primary purpose of the  deployment of the catch workflows was to collect the data reported in the paper and to develop and refine the approaches to removing false positives.
However, the process of determining ground truth on true/false positives inherently involved reaching out to engineers to confirm whether a weak catch was a strong catch.
During the period from mid-September through to the end of October, we reached out to \NumberOfReachouts~engineers, 
with a very simple chat-based message asking whether the observed change was intended or not.
Engineers reported that dismissing an intended change took only a few minutes, and therefore this low bandwidth high signal introduces a very minimal degree of friction into the overall development process, and has negligible effects on developer velocity.

In the process of reaching out to engineers, we were also able to (strongly) catch 8 issues, of which 
4 proved to prevent significant issues that would have led to severe failures had they not been intercepted at diff submission time.
For each of these four cases, we undertook a detailed in-depth analysis to confirm that each met the standard criteria for a severe failure had they not been intercepted and averted by the catching test.

It is  extremely encouraging (from the point of view of return on test effort) 
that from 8 issues prevented, 4 of these proved to be severe. 
This indicates that targeting catch testing at the most sensitive code has a very high return on investment. 
It is extremely unlikely that 8 {\em arbitrary} bugs, chosen it random, would yield 4 that lead to severe errors; 
most bugs are relatively trivial in nature~\cite{mockus:case,zeller:programs}.
The scientific literature~\cite{mockus:case}, as well as public bug-reporting systems, such as Bugzilla, tend to suggest that we should expect approximately 5\%-20\% of all bugs to be serious, whereas our results on this (admittedly limited number of bugs) indicates 43\%.
If these findings scale up, then it would be an extremely encouraging finding concerning  the efficiency of targeting.

At this point in the evolution of our catch workflows, we have too few data points to make strong scientific claims of this nature, but the results are tantalizing glimpses of a world in which we can focus testing on catching bugs, reporting low friction signal to engineers, without them ever having to look at test code itself, let alone write it. 
Most importantly of all, when combined with targeting, it seems that many of the bugs, possibly as much as half, will prove to be severe, thereby focusing test effort where it is most effective.

Of the remaining 4 issues caught, 2 lead to fixes involving landing defensive programming to prevent future regressions, while for the other 2, the diffs were simply abandoned by their authors.
The fact that we were able to catch 7 issues from \NumberOfReachouts~reach outs, gives us an overall true positive rate, as experienced by the engineers who received the signal,  of \OverallEngineerPerceivedTPRate.
When coupled with the almost negligible effects of false positives, this was a very encouraging signal.
We are confident that this baseline performance for engineer-experienced true positive rate can be significantly improved.

\subsection{Determining whether to reach out}
Whether or not either a Human-in-the-Loop or fully automated system should {\em choose} 
to reach out to an engineer is a trade-off between precision and recall, 
and has to take account of the likely friction caused to the diff author, 
who is tasked with checking whether the detected change is expected.
The friction involved is partly related to the simplicity of the changes, and also partly the nature of the test failure.
It is interesting to observe that this is not just an assessment of the likelihood of a true positive, 
but also an assessment of the impact, in terms of friction, caused by the human effort of eliminating a false positive.
The more that can be done to reduce this friction, the lower the precision need be overall, 
and therefore, when trading precision and recall, we can increase recall by effectively reducing the cost of false positives.

To reduce the cost of false positives, we aim to make it exceptionally easy for an engineer to clearly dismiss a signal that merely flags an intended change.
We find that this affects the way the technology should interface to the engineer. 
Rather than showing them a test case and asking them to debug, it is much better, in the first instance, to clearly articulate what the changed behavior 
is (between parent and child) and to simply ask whether this is expected. 
In general, we find that engineers can much more quickly answer this question,  compared to considering test case code.

In the rule based approach, therefore, when we have identified a pattern that likely indicates a true positive, we give it a higher overall score if it additionally would be very quick and easy to dismiss. 
For example, suppose the rule-based pattern that applies would materialize into an engineer reach out with a message like: `this expression, used to evaluate to {\tt true}, but now evaluates to {\tt false}; is that expected?'.
Such a reach out is simple and clear, and if the changed behavior is intended, it is typically easy for the engineer to recognize this.

By contrast, consider a situation where the reach out asks the engineer to confirm whether a change in the order of some subset elements of a large composite data structure is expected.
This will typically take orders of magnitude longer to dismiss, should it prove to be a false positive, compared to the boolean change.
Consequently, we could afford an order of magnitude more false positives per true positive for the `boolean change' style reach out, when compared to the `large composite data structure ordering mismatch' reach out.
In applying the rule based approach to determine true/false positives, we therefore also take into account the ultimate impact on the engineer.

Fortunately, we have found that the catch deployment model significantly reduces the friction caused by false positives.
Firstly, the catch is reported at diff time; a time when the developer has maximum mental context 
on the code change.
This makes the signal highly relevant, which we know is important ~\cite{mhpoh:scam18-keynote}.
Secondly, the initial signal is expressed simply as a `sense-check' on whether a changed behavior is expected or not.
In the case of a false positive, the engineer simply confirms the change is expected, 
which can take a matter of seconds, and we found, in practice, takes no more than a few minutes.
False positives are thus eliminated with alacrity, and consequently a 
higher proportion can be tolerated than were the engineer required to actually look at test  code or code under test.

Finally, only when the sense check fails, 
and the engineer is {\em surprised} by the observed behavior change, 
do they need to consider test case code itself.
Such situations may ultimately prove to be false positives, 
but the fact that the engineer is surprised makes it worth the double-check, even in false positive  situations; 
it helps to improve developer understanding of the code.
This improved understanding is particularly valuable in the rapidly emerging scenario~\cite{xiao:generative,otten:prompting} 
whereby a large amount of code is machine-generated (under engineer supervision).

Overall, therefore, catch workflows can 
gain significant real world impact by preventing serious failures 
and with almost negligible drag on developer velocity, 
and with many other indirect advantages on newly-emerging development processes.

\section{Reducing false positives and identifying likely true positives}
\label{sec:judges}
For any workflow that aims to generate weak catches, whether diff-aware, intent-aware, both or neither, we must ask the question

\begin{quote}
     {\it What is the probability that this weak catch (a test failure) is a strong catch (true positive test failure)?}
\end{quote}

In this section, we report on our initial findings on  false and true positive identification.
We found it more intuitive to combine the false positive and true positive identifications into  a value between -1.0 and 1.0, 
where -1.0 indicates very high probability of a false positive (a strictly weak catch), while 1.0 represents very high probability of a true positive (a strong catch).
We implement this weak/strong catch assessment as a  post-processing phase that is applied to all workflows that attempt to produce weak catches as candidate strong catches.

We call these automated true/false positive decision procedures `assessors'. 
The assessors enable the overall Catching JiTTest approach to (1) filter out false positives and prioritize human-in-the-loop review for high-confidence cases, and (2) manage the precision-recall tradeoff of the overall system.

The assessor workflows are only ever deployed when we have a candidate weak catch.
The existence of a weak catch means that the assessor always has  the tested diff available to it, when forming its judgment.
The assessor, therefore, can always be both diff-aware {\em and} intent-aware, irrespective of whether the workflow that generated the weak catch was, itself, aware of diff- or intent- aware.

We use two different approaches to judging strong catch likelihood; LLM-as-judge \cite{zheng2023judging},  and a more traditional rule-based approach called RubFake: Rule Based False-positive Killing Environment.
These two approaches are complementary. 
While the LLM-as-judge has rich semantics available  to it, its judgments are both nondeterministic and challenging to verify.
We cannot rely on it to give ground truth, but we can aspire to create a rank-based judgment  of a set of weak catches with reasonable rank correlation to the ground truth likelihood of a strong catch.

By contrast, the rule based approach searches for specific patterns that are either present or absent, and makes deterministic verified judgments about whether these patterns are present.
We can rely on the veracity of the rule-based approach in the sense that we know that the claimed patterns are guaranteed to exist, 
but interpreting whether these patterns likely correlate with the strongness of catch also remains more of a question of correlation than of ground truth.

We use these signals together to produce an overall ranking of weak catch candidates, 
with those ranked sufficiently highly being worthy of further consideration.
Ultimately, the aspiration is that this correlation will be found sufficiently close to the ground truth that we will be able to define suitable thresholds, above which our infrastructure will automatically send the outcome  to the diff author.

In our initial  deployment, we also have a human-in-the-loop. 
The purpose of the human is, simultaneously, to determine whether a candidate strong catch signal is sent to the diff author, 
and to provide feedback on true and false positive patterns so that we can iteratively improve both the LLM-as-judge and the rule-based approaches.

\subsection{LLM-as-Judge}
To maximize robustness and reduce the risk of idiosyncratic model errors or biases, 
we employ an ensemble approach, aggregating judgments from several state-of-the-art LLMs 
% NOTE: legal may not want us to name the specific language models. Let's name them explicitly and allow legal to comment on this. I'm flagging here so that we double check with them if they don't flag it.
(including Llama3.3-70B, Gemini 3 Pro, and Claude Sonnet 4). 
Each model is tasked with classifying whether a given weak catch represents an unexpected bug introduced by the code diff.

\subsubsection{Input}
Each LLM-as-Judge receives the following inputs for each candidate test case:
\begin{enumerate}
    \item The code of the weak catch test case
    \item The test execution outcome, including the failure code, message, and test execution trace
    \item The code changes between parent and child (diff)
    \item The inferred diff intent
\end{enumerate}

\subsubsection{Output}
From the execution of the LLM-as-judge ensemble, we obtain  three key outputs for each weak catch case:-

\noindent
{\bf Normalized Token Probability Score (TP Prob)} (-1 to 1): Token probability was only available for Llama models, and we chose Llama3.3-70B and tasked the LLM with a binary classification; determining whether a test failure represented by the weak catch is unexpected. 
    The model outputs a `yes' or `no' response, and the normalized logarithmic probability of the first token (`Yes' or `No')  is assigned to a score between -1 and 1. 
    A `yes' response with high log-probability yields a score close to 1, indicating confidence in a true positive (i.e., strong catch), while a `no' response with high log-probability yields a score close to -1, indicating confidence in a false positive.

\noindent    
{\bf Ensemble Categorical Likelihood Score (Bucket Med)} (-1 to 1): Each LLM  in the ensemble also performs a multi-class classification, producing a categorical output that indicates the likelihood that a weak catch will be unexpected, with possible values of High, Medium, or Low Likelihood. These categorical outputs are subsequently transformed to numerical values, i.e., {`High' : 1.0, `Medium': 0.0, `Low': -1.0}, yielding a scale consistent with the normalized token probability score described above. The final score is determined by taking the median across the ensemble, providing a consensus-based measure of the likelihood that the test failure represents a bug. The median is chosen over the mean because the mapped values are ordinal rather than continuous, ensuring the aggregation method is appropriate for the nature of the data\cite{shepperd:foundations}.

\noindent
{\bf Textual explanation}: The LLM is instructed to provide a brief, structured rationale for its classification. This rationale describes (i) the code behavior change evidenced by the weak catch test failure and (ii) why this behavior change is judged to be unexpected given the diff and its intent. Although the system employs an ensemble of LLM judges, aggregation of these textual rationales is not currently performed. Instead, the ensemble’s value lies in the diversity of explanations, which offers multiple perspectives rather than a synthesized consensus. Each explanation is presented individually, allowing human reviewers to evaluate the performance of each judge, compare their reasoning, and gain insight into the strengths and weaknesses of different approaches.Future deployments may explore methods for aggregating these explanations—such as selecting the rationale from the most confident judge or synthesizing a consensus explanation—to help diff authors more efficiently interpret test failures and determine appropriate follow-up actions.

The LLM-as-Judge ensemble thus produces, for each weak catch, a set of confidence scores and associated rationales, which are surfaced for human review. 
Empirical evaluation on \SpreadsheetWeakCatchesForLLMJudge~weak catches showed that the LLM judges achieved over 98\% precision\footnote
{The tests are drawn from extremely unbalanced population in which almost all of them are false positives. 
Therefore, high precision is expected. 
It could be obtained from a technique that simply assessed every single test as false positive. 
However, our experience was that it was safe to discard those with the lowest scores, and this is what saved 70\% of human effort without, we believe, losing any true positive catches. 
Therefore, the important take-home message from these numbers is the 70\% saving, rather than the 98\% precision.
} 
in identifying false-positive catches when either the normalized token probability score was –1 or the ensemble categorical likelihood score was 0. 
This high precision enabled a reduction of human review workload by approximately 70\%, effectively allowing us to scale evaluation capacity to approximately 4× that of manual review alone.

% In addition, our analysis indicates an opportunity to “shift left” in the pipeline. 
% If the post-generation LLM-as-Judge module can reliably identify test failures that contradict the inferred diff intent, future iterations may avoid generating such weak tests altogether. 
% Designing this earlier-stage filtering mechanism is a key direction for further reducing false positives in the system.

\subsection{Rule-based false positive identification}
Between October and November 2025, we manually analyzed \SpreadsheetWeakCatchesForLLMJudgeManuallyAnalysed~ weak catches, in order to identify patterns of false positive behavior. 
We initially had two engineers performing this process, in which one would conduct the assessment and the other act as a checker on this assessment. 
Subsequently, as we scaled up the deployment, we employed four engineers, each of whom would take a subset of the weak catches and produce an initial assessment which was checked by the others. 

This was an informal process, 
aimed at identifying patterns, 
rather than precisely  computing formally-defined inter-rater agreement or assessment of likelihood of true positive.
However, where we found a good candidate for a strong catch, we  reached out to the author of the tested diff, 
and the diff's reviewers, to check whether they considered it to be a true positive catch. 
This simple  deployment with a Human-in-the-Loop  was aimed at refining the rule base approach. 
Nevertheless, it has already found several important bugs, 
giving us confidence that the overall approach had comfortably passed the `proof of concept' evaluation bar.
Section~\ref{sec:examples-of-bugs}~provides details of the bugs caught and fixed before reaching production 
using this approach, and the kind of failures that were thereby averted.

\subsubsection{False positive  patterns}
Table~\ref{tab:fp-bubfake} presents information about the false positive patterns  found.
We consider three different sources of information when determining a false positive:

\begin{enumerate}
    \item The change made to the code under the test
    \item The generated test code
    \item The execution trace produced when the test is executed on the diff under test
\end{enumerate}

False positive patterns can be found in any or all of these sources of signal.

\begin{table*}

    \scriptsize
    \caption{False positive patterns}
    \label{tab:fp-bubfake}
    \centering
    \begin{tabular}{|l|l||p{9.4cm}|r|}
        \hline
        {\bf Pattern Name}          & {\bf Likelihood} &                     & {\bf Source(s)}    \\
                                    & {\bf of false}   & {\bf  Description}  & {\bf checked}      \\     
                                    & {\bf positive}   & {\bf  of pattern}   & {\bf for pattern}  \\    
         \hline
        
       broken\_test\_runner & high& when the test has failed because the test runner itself failed that this is a sign of an infrastructure issue not a bug detected & execution log \\ \hline
       reflection & high & reflection tends to lead to highly brittle test since it is inherently implementation dependent & execution log, test code\\ \hline
       type\_missmatch & high & often the type of methods changes between parent and child and this can lead to false positive test failure & execution log, test code\\ \hline
       bad\_mock\_smell & medium & where the test appears to be failing due to an incorrect mock, we cannot be sure & execution log\\ \hline
       should\_be\_private\_smell & high & We found that LLMs would generate tests that attempt to enforce private methods& execution log, test code \\ \hline
       method\_must\_be\_protected\_smell & high & We found LLMs would occasionally enforce protective methods & execution log, test code\\ \hline
       mock\_broken & high & we found the LLM's often incorrectly mock & execution log\\ \hline
       data\_provider\_broken & high & we found the LLM's often constructing incorrect data providers & execution log\\ \hline
       not\_implemented\_exception & high & almost certainly the engineer intentionally raised an exception to indicate a method signature has been added without an implementation & execution log \\ \hline
       key\_value\_pair\_change & high & Requiring same key value pairs in the same order is too implementation specific & execution log\\ \hline
       undefined\_variable & high & The undefined variable is likely removed by the code modifications or introduced incorrectly by the test & execution log\\ \hline
       expecting\_particular\_calls\_to\_functions & high & Clearly implementation dependent expectations such as a method being called a certain number of times or with a certain type of parameter & execution log, test code\\ \hline
       web\_server\_down & high & test failed due to infrastructure issues & execution log \\ \hline
       flakiness & low & various patterns that may indicate flaky execution & execution log, test code\\ \hline

    \end{tabular}
\end{table*}

\subsection{Flagging true positive patterns}
Throughout October and November 2025 the Human-in-the-Loop process of analyzing weak catches 
(and consequent occasional reaches out to engineers) gave us insights on true positive patterns as well as false positives.
Given the success of rule-based identification of false positives, we decided to also implement a rule-based approach to identify true positive patterns.
While one can never be 100\% certain that a pattern is a true positive, 
we essentially codified as rules, 
the patterns that we intuitively found ourselves responding to when deciding to reach out to engineers.
These could be regarded as promising patterns for a potential true positive, 
rather than rule based guarantees of a true positive. 

This is to be contrasted with the false positive identification, 
in which any candidate identified as a false positive is essentially eliminated.
Table~\ref{tab:tp-rubfake} presents information on the true positive patterns identified and used.
When codifying true positive likelihood, we took account of the ease with which a wrong 
true positive claim could be eliminated by an engineer with relevant context.

\begin{table*}
    \scriptsize
    \caption{True positive patterns}
    \label{tab:tp-rubfake}
    \centering
    \begin{tabular}{|l|l||p{3.5cm}|p{7.5cm}|}
        \hline
    {\bf Pattern}           & {\bf Execution trace signs}       & {\bf Corroboration in code/intent }        & {\bf Explanation} \\  \hline
        
    Unexpected\_key\_change & key out of bounds                 & No update to keys                          & Indexing into container fails even though the key fields have not been updated \\ \hline
    Empty\_container        & empty container exception         & No changed line directly effects container & A container such as addict, array becomes empty despite not being explicitly changed \\ \hline
    Create\_failure         & object creation exception         & No direct change to object structure              & An exception indicating an object cannot be created is suspicious if the object structure is not updated   \\ \hline
    changed\_bool           & True becomes false or vice versa  & Context dependent                          & When an asserted true becomes false or vice versa service is suspicious, but we need to check whether those have good reason in the code.  \\ \hline
    refactor                & any specific change in behavior  & Intention is to refactor                   & most refactorings should be meaning preserving by construction  \\ \hline
    dead\_code\_removal     &  any specific change in behavior & Intention is dead code removal             & the code removal is expected to be safe and therefore meaning preserving \\ \hline
    null\_value             & value is null                     & Defining expression  not updated           & suspicious when variables not directly implicated in code changes become null  \\ \hline
    monotonic\_change       & existing behavior change         & Intention is only to add new functionality & The change clearly intended only to add behavior. For example, adding logging, but existing behavior changed.\\ \hline
    RBAC                    & assert failure flags RBAC change  & Code does not specifically affect this role & Rule-Based Access Control (RBAC) can involve complex corner cases, and dependences so unexpected changes are Suspicious \\ \hline
    \end{tabular}
\end{table*}

As can be seen, a common theme of the identification of a true positive pattern is that the test identifies a specific kind of change, such as the boolean transitioning from {\tt true} to {\tt false}, thereby causing the test to fail on the child when it passes on the parent, while not seeing any direct attempt to change the boolean-defining expression in the code itself.
We do not claim that it would be prudent to always reach out when such a rule applies, but it certainly adds weight to the overall pool of evidence available, suggesting that it is worth reaching out.

\subsection{Assessing the assessors}
\label{sec:stats}
We analyzed the behavior of our three true/false positive assessors using data from two diff-aware catch workflows (Section~\ref{sec:workflows}), applied to code changes from two distinct groups: `good' diffs (accepted or committed by reviewers) and `bad' diffs (reverted, abandoned, or flagged for changes).
`Good' diffs are those deemed landable by human reviewers, but may still contain bugs. 
`Bad' diffs may be abandoned for various reasons, not necessarily due to bugs. Thus, diff status does not guarantee bug likelihood, but does provide expert-labeled data.

Although these labels are imperfect (`landable' vs. `not landable' rather than `bug-free' vs. `buggy'), they enable cautious statistical analysis. 
After all, were `Good' and `Bad' diffs  to show no differences, it would undermine the value of modern code review, yet empirical studies have repeatedly confirmed its effectiveness since Fagan inspections in the 1970s \cite{bacchelli:expectations,sadowski:modern,fagan:design}.
Given this extensive literature, we expect statistical differences between `Good' and `Bad' diffs and use this expectation to compare test generation methods and assessor performance across these labeled diffs.

Since we are executing multiple statistical tests, there is a probability that some $P$ values that are found to be significant at the $\alpha = 0.05$ level  will be spurious.
Fortunately we are not using the outcomes to make life-altering decisions about drug treatments~\cite{gelman2012why,perneger1998whats,rothman1990no} so we neither want nor need  conservative $p$-value corrections such as Bonferroni or Hochberg~\cite{hochberg:sharper}.
Rather, in common with almost all Software Engineering applications~\cite{mhetal:sbse-tutorial,arcuri:practical} it is not just statistical significance that matters, but the effect size.
Therefore, to cater for the influence of multiple statistical tests, we assess the likelihood of obtaining a statistically significant result 
with a given effect size, using similar data to that we study using a permutation-based resampling procedure~\cite{good2013permutation}.

The resampling computes that the chance of obtaining a single observation of small effect size according to Cohen's $h$ for two samples deemed statistically significantly different according to Fisher's exact test.
That chance is approximately 0.04. 
However, the chance of obtaining a medium or large effect size  for samples of similar size to our `good' and `bad' diffs was computed to be less than $0.00001$.
We thus pay for more head to medium and larger effect sizes (and/or multiple related occurrences of small effect sizes, for which the joint probability of these occurring by chance is low).

%% for latex auto get use --allow-very-weak-duplicates to ensure we screen out same test on same diff with same commit hash.

%% --- generated text from command ... do not edit directly ...
% pasted from appendex
% 

\begin{table*}
\centering
\small
\caption{Inter-rater agreement and rank correlation among assessment methods. Spearman ($\rho$) and Kendall ($\tau$) rank correlations show coefficient if $p < 0.05$, otherwise `--'. Cohen's $\kappa$ for pairwise agreement on FP/TP classification; Fleiss' $\kappa$ and Krippendorff's $\alpha$ for overall agreement. Grayscale intensity indicates strength (darker = stronger). G=Good (Accepted/Closed), B=Bad (other statuses), $G \cup B$=labelled only, UnLab=unlabelled, All=all diffs. Status abbreviations: CLS=Closed, ACC=Accepted, ABD=Abandoned, CHP=Changes Planned, NRS=Needs Revision. Interpretation: Pr=Poor, Sl=Slight, Fr=Fair, Mod=Moderate, Sub=Substantial, AP=Almost Perfect.}
\label{tab:agreement-and-correlation}

\centering
\begin{tabular}{lccccccccccc}
\toprule
\textbf{Spearman Pair Comparison} & \textbf{All} & \textbf{UnLab} & \textbf{$G \cup B$} & \textbf{G} & \textbf{B} & \textbf{CLS} & \textbf{ACC} & \textbf{ABD} & \textbf{CHP} & \textbf{NRS} & \textbf{REV} \\
$n$ (diff count) & 531 & 275 & 256 & 150 & 106 & 90 & 60 & 77 & 12 & 16 & 1 \\
\midrule
RubFake vs TPP & \cellcolor[gray]{0.85}-0.146 & -- & \cellcolor[gray]{0.80}-0.196 & \cellcolor[gray]{0.77}-0.226 & -- & -- & \cellcolor[gray]{0.73}-0.265 & -- & -- & -- & -- \\
RubFake vs Bucket & -- & -- & -- & -- & -- & -- & -- & -- & -- & \cellcolor[gray]{0.47}\textcolor{white}{\textbf{+0.535}} & -- \\
TPP vs Bucket & \cellcolor[gray]{0.74}+0.260 & \cellcolor[gray]{0.73}+0.266 & \cellcolor[gray]{0.75}+0.254 & \cellcolor[gray]{0.75}+0.249 & \cellcolor[gray]{0.77}+0.229 & \cellcolor[gray]{0.69}+0.313 & -- & -- & -- & \cellcolor[gray]{0.49}\textcolor{white}{\textbf{+0.512}} & -- \\
\bottomrule
\end{tabular}

\vspace{1em}

\centering
\begin{tabular}{lccccccccccc}
\toprule
\textbf{Kendall Pair Comparison} & \textbf{All} & \textbf{UnLab} & \textbf{$G \cup B$} & \textbf{G} & \textbf{B} & \textbf{CLS} & \textbf{ACC} & \textbf{ABD} & \textbf{CHP} & \textbf{NRS} & \textbf{REV} \\
$n$ (diff count) & 531 & 275 & 256 & 150 & 106 & 90 & 60 & 77 & 12 & 16 & 1 \\
\midrule
RubFake vs TPP & \cellcolor[gray]{0.89}-0.113 & -- & \cellcolor[gray]{0.85}-0.152 & \cellcolor[gray]{0.82}-0.182 & -- & \cellcolor[gray]{0.84}-0.164 & \cellcolor[gray]{0.78}-0.217 & -- & -- & -- & -- \\
RubFake vs Bucket & -- & -- & -- & -- & -- & -- & -- & -- & -- & \cellcolor[gray]{0.53}+0.469 & -- \\
TPP vs Bucket & \cellcolor[gray]{0.79}+0.212 & \cellcolor[gray]{0.78}+0.219 & \cellcolor[gray]{0.80}+0.205 & \cellcolor[gray]{0.80}+0.204 & \cellcolor[gray]{0.82}+0.183 & \cellcolor[gray]{0.74}+0.259 & -- & -- & -- & \cellcolor[gray]{0.56}+0.440 & -- \\
\bottomrule
\end{tabular}

\vspace{1em}

\begin{tabular}{l|cccccccccc}
\toprule
False Positive Pair & All & UnLab & $G \cup B$ & G & B & CLS & ACC & ABD & CHP & NRS \\
\midrule
Rubfake vs TP Prob & \cellcolor[gray]{1.00}-0.07 (Pr) & \cellcolor[gray]{1.00}-0.01 (Pr) & \cellcolor[gray]{1.00}-0.15 (Pr) & \cellcolor[gray]{1.00}-0.17 (Pr) & \cellcolor[gray]{1.00}-0.12 (Pr) & \cellcolor[gray]{1.00}-0.16 (Pr) & \cellcolor[gray]{1.00}-0.17 (Pr) & \cellcolor[gray]{1.00}-0.10 (Pr) & \cellcolor[gray]{1.00}-0.24 (Pr) & \cellcolor[gray]{1.00}-0.12 (Pr) \\
Rubfake vs Bucket Med & \cellcolor[gray]{1.00}-0.07 (Pr) & \cellcolor[gray]{1.00}-0.09 (Pr) & \cellcolor[gray]{1.00}-0.06 (Pr) & \cellcolor[gray]{1.00}-0.04 (Pr) & \cellcolor[gray]{1.00}-0.07 (Pr) & \cellcolor[gray]{1.00}-0.04 (Pr) & \cellcolor[gray]{1.00}-0.05 (Pr) & \cellcolor[gray]{1.00}-0.16 (Pr) & \cellcolor[gray]{0.88}0.12 (Sl) & \cellcolor[gray]{0.74}0.26 (Fr) \\
TP Prob vs Bucket Med & \cellcolor[gray]{0.80}0.20 (Sl) & \cellcolor[gray]{0.80}0.20 (Sl) & \cellcolor[gray]{0.80}0.20 (Fr) & \cellcolor[gray]{0.82}0.18 (Sl) & \cellcolor[gray]{0.80}0.20 (Sl) & \cellcolor[gray]{0.77}0.23 (Fr) & \cellcolor[gray]{0.90}0.10 (Sl) & \cellcolor[gray]{0.86}0.14 (Sl) & \cellcolor[gray]{1.00}-0.03 (Pr) & \cellcolor[gray]{0.38}\textcolor{white}{\textbf{0.62 (Sub)}} \\
\midrule
Fleiss' $\kappa$ (all 3) & \cellcolor[gray]{0.99}0.01 (Sl) & \cellcolor[gray]{0.97}0.03 (Sl) & \cellcolor[gray]{1.00}-0.02 (Pr) & \cellcolor[gray]{1.00}-0.04 (Pr) & \cellcolor[gray]{1.00}-0.01 (Pr) & \cellcolor[gray]{1.00}-0.03 (Pr) & \cellcolor[gray]{1.00}-0.06 (Pr) & \cellcolor[gray]{1.00}-0.06 (Pr) & \cellcolor[gray]{1.00}-0.05 (Pr) & \cellcolor[gray]{0.75}0.25 (Fr) \\
Krippendorff's $\alpha$ (all 3) & \cellcolor[gray]{0.99}0.01 (Sl) & \cellcolor[gray]{0.97}0.03 (Sl) & \cellcolor[gray]{1.00}-0.02 (Pr) & \cellcolor[gray]{1.00}-0.04 (Pr) & \cellcolor[gray]{1.00}-0.01 (Pr) & \cellcolor[gray]{1.00}-0.02 (Pr) & \cellcolor[gray]{1.00}-0.06 (Pr) & \cellcolor[gray]{1.00}-0.05 (Pr) & \cellcolor[gray]{1.00}-0.02 (Pr) & \cellcolor[gray]{0.73}0.27 (Fr) \\
\midrule
Sample size (n) & 531 & 275 & 256 & 150 & 106 & 90 & 60 & 77 & 12 & 16 \\
\bottomrule
\end{tabular}

\vspace{1em}

\begin{tabular}{l|cccccccccc}
\toprule
True Positive Pair & All & UnLab & $G \cup B$ & G & B & CLS & ACC & ABD & CHP & NRS \\
\midrule
Rubfake vs TP Prob & \cellcolor[gray]{1.00}-0.00 (Pr) & \cellcolor[gray]{1.00}-0.02 (Pr) & \cellcolor[gray]{0.98}0.02 (Sl) & \cellcolor[gray]{0.99}0.01 (Sl) & \cellcolor[gray]{0.99}0.01 (Sl) & \cellcolor[gray]{0.99}0.01 (Sl) & \cellcolor[gray]{0.97}0.03 (Sl) & \cellcolor[gray]{1.00}-0.04 (Pr) & \cellcolor[gray]{1.00}-0.16 (Pr) & \cellcolor[gray]{0.62}0.38 (Fr) \\
Rubfake vs Bucket Med & \cellcolor[gray]{1.00}-0.02 (Pr) & \cellcolor[gray]{1.00}-0.03 (Pr) & \cellcolor[gray]{1.00}-0.01 (Pr) & \cellcolor[gray]{0.93}0.07 (Sl) & \cellcolor[gray]{1.00}-0.10 (Pr) & \cellcolor[gray]{1.00}0.00 (Sl) & \cellcolor[gray]{0.76}0.24 (Fr) & \cellcolor[gray]{1.00}-0.16 (Pr) & \cellcolor[gray]{1.00}-0.12 (Pr) & \cellcolor[gray]{0.82}0.18 (Sl) \\
TP Prob vs Bucket Med & \cellcolor[gray]{0.88}0.12 (Sl) & \cellcolor[gray]{0.86}0.14 (Sl) & \cellcolor[gray]{0.90}0.10 (Sl) & \cellcolor[gray]{0.93}0.07 (Sl) & \cellcolor[gray]{0.90}0.10 (Sl) & \cellcolor[gray]{0.91}0.09 (Sl) & \cellcolor[gray]{0.93}0.07 (Sl) & \cellcolor[gray]{0.91}0.09 (Sl) & \cellcolor[gray]{0.94}0.06 (Sl) & \cellcolor[gray]{0.88}0.12 (Sl) \\
\midrule
Fleiss' $\kappa$ (all 3) & \cellcolor[gray]{1.00}-0.02 (Pr) & \cellcolor[gray]{1.00}-0.02 (Pr) & \cellcolor[gray]{1.00}-0.02 (Pr) & \cellcolor[gray]{1.00}-0.03 (Pr) & \cellcolor[gray]{1.00}-0.05 (Pr) & \cellcolor[gray]{1.00}-0.01 (Pr) & \cellcolor[gray]{1.00}-0.05 (Pr) & \cellcolor[gray]{1.00}-0.10 (Pr) & \cellcolor[gray]{1.00}-0.13 (Pr) & \cellcolor[gray]{0.81}0.19 (Sl) \\
Krippendorff's $\alpha$ (all 3) & \cellcolor[gray]{1.00}-0.02 (Pr) & \cellcolor[gray]{1.00}-0.02 (Pr) & \cellcolor[gray]{1.00}-0.02 (Pr) & \cellcolor[gray]{1.00}-0.03 (Pr) & \cellcolor[gray]{1.00}-0.05 (Pr) & \cellcolor[gray]{1.00}-0.01 (Pr) & \cellcolor[gray]{1.00}-0.05 (Pr) & \cellcolor[gray]{1.00}-0.10 (Pr) & \cellcolor[gray]{1.00}-0.09 (Pr) & \cellcolor[gray]{0.79}0.21 (Fr) \\
\midrule
Sample size (n) & 531 & 275 & 256 & 150 & 106 & 90 & 60 & 77 & 12 & 16 \\
\bottomrule
\end{tabular}

\vspace{0.5em}
\begin{minipage}{\linewidth}
\footnotesize
\textit{Assessment methods:} RubFake = rule-based approach; TPP = LLM log-likelihood probability; Bucket = bucketed LLM-ensemble assessment.
\end{minipage}
\end{table*}

\begin{table*}
\centering
\caption{Fisher's Exact Test p-values comparing assessment rates between good and bad diffs for each catch approach. True Positive (TP) columns show p-values for score $> 0$; False Positive (FP) columns show p-values for score $< 0$. For significant results, effect size (Cohen's $h$, \cite{cohen:statistical-power}) and direction are shown: N=negligible ($|h|<0.2$), S=small ($0.2 \leq |h|<0.5$), M=medium ($0.5 \leq |h|<0.8$), L=large ($|h| \geq 0.8$); G=good has higher rate, B=bad has higher rate. Significant p-values ($p < 0.05$) are shown in bold. P-values are uncorrected for multiple testing.}
\label{tab:approach-pvalues}
\begin{tabular}{lrrr|rrr}
\toprule
 & \multicolumn{3}{c}{True Positive (score $> 0$)} & \multicolumn{3}{c}{False Positive (score $< 0$)} \\
Approach & Rubfake & TP Prob & Bucket Med & Rubfake & TP Prob & Bucket Med \\
\midrule
intent-aware workflow & 0.586 & 0.094 & 0.104 & 0.736 & 0.094 & \textbf{0.003 (S,G)} \\
dodgy diff workflow & \textbf{0.002 (S,B)} & \textbf{$<$0.001 (M,B)} & \textbf{$<$0.001 (M,B)} & 0.237 & \textbf{$<$0.001 (M,G)} & \textbf{0.002 (S,G)} \\
\bottomrule
\end{tabular}
\end{table*}

%%%%
%
% ------------------------------------- end of generated text
%

Table~\ref{tab:approach-pvalues} shows the results for true and false positive assessment according to the three different ways of determining whether a test failure is true or false positive, for each of the two different test generation techniques.
As can be seen, the true positive assessor for all three has a significantly higher rate in the diffs labeled as `bad', compared to those labeled as `good' when using the simple `dodgy diff' test generation approach.
Similarly, the two LLM-based assessors have a noticeably higher false positive assessment rate for tests generated on the good diffs.

Also, Table~\ref{tab:agreement-and-correlation} reveals modest levels of inter-rater agreement and rank correlation ($\rho \le \highestRankCorrlection$), especially for the `Needs Revisions' sub-category of diffs.
Based on this overall relatively modest correlation, one might be tempted to assume that some/all assessors simply add noise, but low correlation {\em could} also indicate complementarity. 
This complementarity interpretation becomes much {\em more probable} given the discriminating power revealed by the Fisher exact test, and results showing the extremely low probability of achieving these effect sizes purely by chance.

Therefore, we conclude that the assessors have some true/false discriminatory power, and that they are likely complementary.
However, there remains no guarantee that, even optimally combined, they cannot be significantly improved upon in terms of ability to correctly predict true/false positive. 
We leave future optimization as an open problem for further research. Our results certainly do indicate that such further research would likely  bear significant fruit and industrial impact.

\section {Coincidental Hardening tests }

% and time >= 1762819200 --- 11th Nov 2025
% ---- BOTH catch approaches and time >= 1762819200 --- 11th Nov 2025
% -- all diffs last month attempted with catch workflows from chronos             --- 2729 diffs --- 62 tests max  8797 tests 
% -- AND outcome='Passed'  -- hardening                                           --- 2517 diffs --- 62 tests max  7987 hardening
% -- AND outcome='Failed'   --weak catches                                        ---  282 diffs ---  9 tests max   405 weak catching

% --- all tests 
% -- and catch_approach_used = 'make_and_kill_mutants_from_diff_intent_and_risks' --- 1471 diffs --- 42 tests max  2923 tests
% -- and catch_approach_used = 'make_tests_from_parent_and_dodgy_diff'            --- 1982 diffs --- 62 tests max  5880 tests

% --- all passing tests 
% -- and catch_approach_used = 'make_and_kill_mutants_from_diff_intent_and_risks' --- 1271 diffs --- 42 tests max   2551 passing tests
% -- and catch_approach_used = 'make_tests_from_parent_and_dodgy_diff'            --- 1763 diffs --- 62 tests max   5436 passing tests

% --- all failing  tests 
% -- and catch_approach_used = 'make_and_kill_mutants_from_diff_intent_and_risks' ---  168 diffs ---  5 tests max    200 failing tests
% -- and catch_approach_used = 'make_tests_from_parent_and_dodgy_diff'            ---  140 diffs ---  9 tests max    205 failing tests

Generating tests to find catches can be expensive.
It would also be inefficient if that were all we did with the workflow, because most code changes are bug free, and therefore any attempt to generate tests to catch bugs in them would be futile.
Fortunately, hardening tests are a very natural and useful {\em byproduct} of the workflows to generate catching tests.
Table~\ref{tab:coincidental-harden} shows the results for coincidentally hardening tests found during attempts to generate weak catching tests in the period from 11th November 202 to 11th December 2025.

Although coincidental catching tests generated from hardening are rare (see Table~\ref{tab:numbers-of-weak-catch-results}), 
coincidental hardening test generated by catching are highly prevalent, as Table~\ref{tab:coincidental-harden} shows.
Therefore, one natural way to approach the overall combined problems of hardening and catching test generation, is to simultaneously do both: primarily focussing on generating catching tests, while also harvesting hardening test generated during this process.

\begin{table}
\centering
\begin{tabular}{|l|l|r|r|r|}
\hline
\textbf{}         & \textbf{}        & \textbf{}       & \textbf{Max}   & \textbf{Total} \\
\textbf{}         & \textbf{}        & \textbf{No.}    & \textbf{Tests} & \textbf{No.} \\
\textbf{Catch }   & \textbf{}        & \textbf{of}     & \textbf{Per}   & \textbf{of} \\
\textbf{Approach} & \textbf{Outcome} & \textbf{Diffs}  & \textbf{Diff}  & \textbf{Tests} \\
% \hline
% \multicolumn{5}{|c|}{\textbf{All Diffs (last month, catch workflows from Chronos, time $\geq$ 11 Nov 2025)}} \\
\hline
\hline
\multicolumn{5}{|c|}{\textbf{All Tests}} \\
\hline
intent-aware workflow & All & 1471 & 42 & 2923 \\
dodgy diff workflow             & All & 1982 & 62 & 5880 \\
\hline
Both approaches & All        & 2729 & 62 & 8797 \\
\hline
\multicolumn{5}{|c|}{\textbf{All Passing Tests}} \\
\hline
intent-aware workflow & Passed & 1271 & 42 & 2551 \\
dodgy diff workflow             & Passed & 1763 & 62 & 5436 \\
\hline
Both approaches & Passed     & 2517 & 62 & 7987 \\
\hline
\multicolumn{5}{|c|}{\textbf{All Failing Tests}} \\
\hline
intent-aware workflow & Failed & 168 & 5 & 200 \\
dodgy diff workflow             & Failed & 140 & 9 & 205 \\
\hline
Both approaches & Failed     &  282 &  9 &  405 \\
\hline
\hline
\end{tabular}
\caption{Co-incidental Hardening from Catch Workflows}
\label{tab:coincidental-harden}
\end{table}

\section {Related work}
Automated test generation has a long history, dating back to the 1960s~\cite{boyer:select}, and remaining active through the decades since \cite{korel90_dynamic}.
Currently, established test generation techniques tend to use either
search-based and evolutionary approaches~\cite{mcminn:survey,fraser:evosuite,fraser:sf110,arcuri:practical}
or Symbolic execution~\cite{godefroid:dart,sen:cute,cadar:three-decades}.
Research has led to the introduction of practical publicly available test generation tools such as EvoSuite~\cite{fraser:evosuite}, 
Randoop~\cite{pacheco:randoop} and Klee~\cite{cadar:klee}.
More recent studies have focussed on LLM-based code generators  such as Codex, Copilot, and Code Llama~\cite{mhetal:fse24-llm,schafer2023empirical,liu2024llm,chen2024chatunitest,wang:llm-test-survey,ryan2024code}.

Existing tools primarily generate unit tests to improve coverage, but coverage may not entail fault revelation; recent research~\cite{mike:icse17} shows mutation testing is more effective, leading to its practical adoption at Meta~\cite{foster:mutation}. 
Whether mutation- or coverage-guided, these approaches automate human-style unit test writing~\cite{mhetal:fse24-llm,anand2012automated,fraser:evosuite,mcminn:past-present-and-future,mh:icst15-keynote,ryan2024code,chen2024chatunitest,schafer2023empirical,liu2024llm}, thereby focusing  on hardening rather than catching test generation.

By contrast, the focus of this paper is on catching unit tests which are very different: they {\em fail} at generation time,
a property they share with techniques such as fuzzing~\cite{manes:fuzzing} and system level test generation with tools like Sapienz~\cite{mao:sapienz:16,mhetal:ssbse18-keynote}.
However, unlike these approaches to catching bugs, we focus on generating catching tests with respect to a general oracle rather than
the implicit oracle ~\cite{ebetal:oracle}.

We believe that the `Catching JiTTest' challenge~\cite{mhpohss:harden} will become increasingly important, due to increasing velocity of software development and consequent volume of code changes and concomitant issues~\cite{cui2024Productivity}.
Tackling this challenge requires 
determining whether a test failure is a true positive, relating directly to the Oracle Problem\cite{ebetal:oracle}. 
Rule-based approaches, such as those leveraging static and dynamic program analysis~\cite{lakhotia:rule-based,zhang:empirically}, have been proposed to filter out flaky or irrelevant failures. 
Recent work has explored the use of large language models (LLMs) for classifying test failures and improving oracle quality~\cite{molina:test,hossain:togll,ibrahimzada2022perfect}, 
as well as for automating the assessment and improvement of test oracles~\cite{ebetal:oracle,gjetal:Empirical-oracle}.

 Techniques like coverage-guided prompting~\cite{pizzorno:coverup,ryan2024code,mhetal:fse24-llm}, adaptive test generation~\cite{schafer_adaptive_2023}, and prompt engineering~\cite{white_chatgpt_2023,shin2023prompt} are not focused on Catching JiTTesting.
 Nevertheless, these methods can enhance relevance to code changes, increasing the likelihood of generating strong catches.

In the past decade, Meta has made considerable investment in automated unit test generation using
a variety of testing techniques including
Metamorphic Testing~\cite{jaetal:mia},
Mutation testing~\cite{foster:mutation},
Observation based testing~\cite{mhetal:TestGen-obs}, and
LLM based testing~\cite{mhetal:fse24-llm}.
These unit based techniques are complemented by automated end-to-end testing~\cite{kmetal:fausta,kinga:enhancing}, and social level testing \cite{jaetal:gi20-keynote,jaetal:mia,tuli:simulation,jaetal:ease21-keynote}, and  static analyses ~\cite{distefano:scaling}, such as
Infer~\cite{movefast} and Zoncolan.
However, this is the first approach at Meta 
to focus on `catching' rather than `hardening'.

% to work in 

% "Most defects are minor and do not result in catastrophic failures."
% — [Mockus et al., 2000]
% "A small number of defects are responsible for the majority of failures experienced by users."
% — [Anderson et al., 2016]
% "Testing is effective at finding simple, shallow bugs, but less so at finding deep, complex bugs that may cause serious failures."
% — [Basili & Selby, 1987]

\section{Conclusions and Future Work}
We reported experience and empirical results from mutation-guided, LLM-based test generation targeting risky code changes (diffs) to uncover faults just before production—the Catching JiTTest Challenge~\cite{mhpohss:harden}. 
This approach has prevented severe bugs from landing at Meta. 
Diff and intent awareness greatly increase the number of weak catches found. 
To transition from weak to strong catches, LLM-based and rule-based true/false positive assessors reduce human review load by 70\%, with statistical analysis showing strong agreement with human labels.

A key advantage is that engineers rarely need to write or review tests, avoiding AI code quality issues~\cite{fu2025securityweaknessescopilotgeneratedcode,gitclear:quality}. 
Tests are generated and assessed automatically; engineers only confirm if changes are intended, with negligible impact on development velocity. 
When changes are unexpected, engineers use test code to quickly identify and fix issues.
We hope these results encourage further research on just-in-time catching test generation and provide scalable, industrially relevant baseline results for future improvements.

\section{ACKNOWLEDGMENTS}
We  would like to thank the leadership of Meta's Product Compliance and Privacy, Fundamental Artificial Intelligence Research (FAIR), Developer Infrastructure (DevInfra), and Instagram Product Foundation teams for supporting our work over the past decade, and the many Meta software engineers and testers whose experience, expertise, and engagement have helped  shape the ideas presented here. 
We would also like to thank the many academics and other researchers with whom we have had the enormous pleasure to interact over 3+ decades of research work on verification, testing and AI.

\newpage
% \balance

%
% The next two lines define the bibliography style to be used, and the bibliography file.
\bibliographystyle{ACM-Reference-Format}
\bibliography{slice}

@string{acm = "Association for Computer Machinery"}

@string{fse = "Foundations of Software Engineering"}

@string{ieee = "Institute of Electric and Electronic Engineers"}

@string{pldi = "ACM SIGPLAN Conference on Programming Language Design
                and Implementation"}

@string{lncs = "Springer Lecture Notes in Computer Science"}

@string{sigplan = "ACM SIGPLan Notices"}

@string{stvr = "Software Testing, Verification and Reliability"}

@misc{abreu:DRS,
      title={Moving Faster and Reducing Risk: {U}sing {LLMs} in Release Deployment}, 
      author={Rui Abreu and Vijayaraghavan Murali and Peter C Rigby and Chandra Maddila and Weiyan Sun and Jun Ge and Kaavya Chinniah and Audris Mockus and Megh Mehta and Nachiappan Nagappan},
      year={2024},
      eprint={2410.06351},
      archivePrefix={arXiv},
      primaryClass={cs.SE},
      url={https://arxiv.org/abs/2410.06351}, 
}

@inproceedings{anand2012automated,
  title={Automated concolic testing of smartphone apps},
  author={Anand, Saswat and Naik, Mayur and Harrold, Mary Jean and Yang, Hongseok},
  booktitle={Proceedings of the ACM SIGSOFT 20th International Symposium on the Foundations of Software Engineering},
  pages={59},
  year={2012},
  organization={ACM}
}

@InProceedings{arcuri:practical,
author = "Andrea Arcuri and Lionel Briand",
title = "A Practical Guide for Using Statistical Tests to Assess Randomized Algorithms in Software Engineering",
 booktitle = {$33^{rd}$ International Conference on Software Engineering ({ICSE'11})},
 year = {2011},
 isbn = {978-1-4503-0445-0},
 location = {Waikiki, Honolulu, HI, USA},
 pages = {1--10},
 numpages = {10},
 publisher = {ACM},
 address = {New York, NY, USA}
}

@inproceedings{bacchelli:expectations,
  title={Expectations, outcomes, and challenges of modern code review},
  author={Bacchelli, Alberto and Bird, Christian},
  booktitle={2013 35th International Conference on Software Engineering (ICSE)},
  pages={712--721},
  year={2013},
  organization={IEEE}
}

@InProceedings{boyer:select,
author = "Robert S. Boyer and Bernard Elspas and Karl N. Levitt",
title = {{SELECT} -- a Formal System for Testing and Debugging Programs by Symbolic Execution},
 booktitle = {International Conference on Reliable Software},
 year = {1975},
 location = {Los Angeles, California},
 pages = {234--245},
 numpages = {12},
 publisher = {ACM},
 address = {New York, NY, USA}
}

@inproceedings{cadar:klee,
  title={Klee: unassisted and automatic generation of high-coverage tests for complex systems programs.},
  author={Cadar, Cristian and Dunbar, Daniel and Engler, Dawson R and others},
  booktitle={OSDI},
  volume={8},
  pages={209--224},
  year={2008}
}

@article{cadar:three-decades,
 author = {Cristian Cadar and Koushik Sen},
 title = {Symbolic Execution for Software Testing: Three Decades Later},
 journal = {Communications of the ACM},
 issue_date = {February 2013},
 volume = {56},
 number = {2},
 month = feb,
 year = {2013},
 issn = {0001-0782},
 pages = {82--90},
 numpages = {9},
 publisher = {ACM},
 address = {New York, NY, USA}
}

@book{cohen:statistical-power,
  author        = "Jacob Cohen",
  title         = "Statistical Power Analysis for the Behavioral Sciences (second ed.)",
  publisher     = "Lawrence Erlbaum Associates",
  address       = "New Jersey",
  year          = "1988"
}

@article{cui2024Productivity,
	author = {Cui, Kevin Zheyuan and Demirer, Mert and Jaffe, Sonia and Musolff, Leon and Peng, Sida and Salz, Tobias},
	journal = {An {MIT} Exploration of Generative {AI}},
	year = {2024},
	month = {mar 27},
	note = {https://mit-genai.pubpub.org/pub/v5iixksv},
	publisher = {MIT},
	title = {The {Productivity} {Effects} of {Generative} {AI}: Evidence from a {Field} {Experiment} with {GitHub} {Copilot}},
}

@article{distefano:scaling,
  title={Scaling static analyses at Facebook},
  author={Distefano, Dino and F{\"a}hndrich, Manuel and Logozzo, Francesco and O'Hearn, Peter W},
  journal={Communications of the ACM},
  volume={62},
  number={8},
  pages={62--70},
  year={2019},
  publisher={ACM New York, NY, USA}
}

@Article{fagan:design,
  author =	"Michael E. Fagan",
  title =	"Design and code inspections to reduce errors in code development",
  journal =	"{IBM} Systems Journal",
  volume =	"15",
  number =	"3",
  pages =	"182--211",
  year = 	"1976",
}

@Article{feitelson:deployment,
  title =	"Development and Deployment at {F}acebook",
  author =	"Dror G. Feitelson and Eitan Frachtenberg and Kent L. Beck",
  journal =	"{IEEE} Internet Computing",
  year = 	"2013",
  number =	"4",
  volume =	"17",
  pages =	"8--17"
}

@InProceedings{fraser:evosuite,
  title =	"{EvoSuite}: automatic test suite generation for object-oriented software",
  author =	"Gordon Fraser and Andrea Arcuri",
  booktitle = "$8^{th}$ European Software Engineering Conference and the {ACM SIGSOFT} Symposium on the Foundations of Software Engineering ({ESEC/FSE '11})",
  month = "September 5th - 9th",
  publisher =	"ACM",
  year = 	"2011",
  ISBN = 	"978-1-4503-0443-6",
  pages =	"416--419"
}

@article{fraser:sf110,
  title={A large-scale evaluation of automated unit test generation using evosuite},
  author={Fraser, Gordon and Arcuri, Andrea},
  journal={ACM Transactions on Software Engineering and Methodology (TOSEM)},
  volume={24},
  number={2},
  pages={1--42},
  year={2014},
  publisher={ACM New York, NY, USA}
}

@misc{fu2025securityweaknessescopilotgeneratedcode,
      title={Security Weaknesses of {Copilot}-Generated Code in {GitHub} Projects: An Empirical Study}, 
      author={Yujia Fu and Peng Liang and Amjed Tahir and Zengyang Li and Mojtaba Shahin and Jiaxin Yu and Jinfu Chen},
      year={2025},
      eprint={2310.02059},
      archivePrefix={arXiv},
      primaryClass={cs.SE},
      url={https://arxiv.org/abs/2310.02059}, 
}

@article{gelman2012why,
  title={Why we (usually) don't have to worry about multiple comparisons},
  author={Gelman, Andrew and Hill, Jennifer and Yajima, Masanao},
  journal={Journal of research on educational effectiveness},
  volume={5},
  number={2},
  pages={189--211},
  year={2012},
  publisher={Taylor \& Francis}
}

@misc{gitclear:quality,
author = "Gitclear",
url = "https://www.gitclear.com/coding_on_copilot_data_shows_ais_downward_pressure_on_code_quality",
title = "Coding on {Copilot}: 2023 Data Suggests Downward Pressure on Code Quality"
}

@InProceedings{godefroid:dart,
  title =	"{DART}: directed automated random testing",
  author =	"Patrice Godefroid and Nils Klarlund and Koushik Sen",
  booktitle =	"Programming Language Design and Implementation ({PLDI 2005})",
  publisher =	"ACM",
  year = 	"2005",
  editor =	"Vivek Sarkar and Mary W. Hall",
  ISBN = 	"1-59593-056-6",
  pages =	"213--223"}

@inproceedings{goffi:automatic,
  title={Automatic generation of oracles for exceptional behaviors},
  author={Goffi, Alberto and Gorla, Alessandra and Ernst, Michael D and Pezz{\`e}, Mauro},
  booktitle={Proceedings of the 25th international symposium on software testing and analysis},
  pages={213--224},
  year={2016}
}

@book{good2013permutation,
  title={Permutation tests: a practical guide to resampling methods for testing hypotheses},
  author={Good, Phillip},
  year={2013},
  publisher={Springer Science \& Business Media}
}

@article{gjetal:Empirical-oracle,
  author    = {Gunel Jahangirova and
               David Clark and
               Mark Harman and
               Paolo Tonella},
  title     = {An Empirical Validation of Oracle Improvement},
  journal   = {{IEEE} Transactions on Software Engineering},
  volume    = {47},
  number    = {8},
  pages     = {1708--1728},
  year      = {2021},
  url       = {https://doi.org/10.1109/TSE.2019.2934409},
  doi       = {10.1109/TSE.2019.2934409},
  bibsource = {dblp computer science bibliography, https://dblp.org}
}

@article{ebetal:oracle,
title = "The Oracle Problem in Software Testing: {A} Survey",
author = "Earl T. Barr and Mark Harman and Phil McMinn and Muzammil Shahbaz and Shin Yoo",
journal = "{IEEE} Transactions on Software Engineering",
year={2015}, 
month={May}, 
volume={41}, 
number={5}, 
pages={507-525}
}

@article{yjmh:analysis,
author = "Yue Jia and Mark Harman",
title = "An Analysis and Survey of the Development of Mutation Testing", 
journal =  "{IEEE} Transactions on Software Engineering",
month = "September--October",
year = 2011, 
Volume =  37, 
number = 5, 
pages = "649 -- 678" 
}

@incollection{mhetal:sbse-tutorial,
author = "Mark Harman and Phil McMinn and {Jerffeson Teixeira de} Souza and Shin Yoo",
title = "Search Based Software Engineering: {T}echniques, Taxonomy, Tutorial",
booktitle = "Empirical software engineering and verification: {LASER 2009-2010}",
publisher = "Springer",
note = "{LNCS 7007}",
pages = "1--59",
year   = "2012",
editor = "Bertrand Meyer and Martin Nordio"
}

@inproceedings{mhetal:LLM-survey,
    author = "Angela Fan and Beliz Gokkaya and Mitya Lyubarskiy and Mark Harman and Shubho Sengupta and Shin Yoo and Jie Zhang",
    title = "{L}arge {L}anguage {M}odels for {S}oftware {E}ngineering: {S}urvey and Open Problems",
    booktitle = "{ICSE} {F}uture of {S}oftware {E}ngineering ({FoSE} 2023)" ,
    year = 2023,
}

@inproceedings{jaetal:ease21-keynote,
author = "John Ahlgren and Kinga Bojarczuk and Sophia Drossopoulou and Inna Dvortsova and Johann George and Natalija Gucevska and Mark Harman and Maria Lomeli and Simon Lucas and Erik Meijer and Steve Omohundro and Rubmary Rojas and Silvia Sapora and Jie M. Zhang and Norm Zhou",
title = "Facebook's Cyber--Cyber and Cyber--Physical Digital Twins (keynote paper)",
booktitle ="25th International Conference on Evaluation and Assessment in Software Engineering ({EASE 2021})",
month = "June",
year = 2021,
address = "Virtual",
}

@InProceedings{jaetal:gi20-keynote,
  author =	"John Ahlgren and Maria Eugenia Berezin and Kinga
		 Bojarczuk and Elena Dulskyte and Inna Dvortsova and
		 Johann George and Natalija Gucevska and Mark Harman and
		 Ralf Laemmel and Erik Meijer and Silvia Sapora and
		 Justin Spahr-Summers",
  title =	"{WES}: Agent-based User Interaction Simulation on Real
		 Infrastructure (keynote paper)",
  booktitle =	"$8^{th}$ Genetic improvement workshop ({GI at ICSE} 2020)",
  year = 	"2020",
  month =	"3 " # jul,
  editor =	"Shin Yoo and Justyna Petke and Westley Weimer and
		 Bobby R. Bruce",
  publisher =	"ACM",
  pages =	"276--284",
}

@inproceedings{mhetal:ssbse18-keynote,
author = "Nadia Alshahwan and Xinbo Gao and Mark Harman and Yue Jia and Ke Mao and Alexander Mols and Taijin Tei and Ilya Zorin",
title = "Deploying Search Based Software Engineering with {S}apienz at {F}acebook (keynote paper)",
booktitle = "$10^{th}$ International Symposium  on Search Based Software Engineering ({SSBSE 2018})",
address = "{M}ontpellier, {F}rance", 
month = "{S}eptember 8th-10th",
  pages     = {3--45},
  year      = {2018},
  note = "Springer {LNCS} 11036",
  editors = "Thelma Colanzi and Phil McMinn"
}

@inproceedings{mhpoh:scam18-keynote,
author = "Mark Harman and Peter {O'H}earn",
title = "From Start-ups to Scale-ups: {O}pportunities and Open Problems for Static and Dynamic Program Analysis (keynote paper)",
booktitle = "$18^{th}$ {IEEE} International Working Conference on Source Code Analysis and Manipulation ({SCAM 2018})",
address = "{M}adrid, {S}pain",
month = "{S}eptember 23rd-24th",
  year      = {2018},
  pages = "1--23"
}

@InProceedings{mh:icst15-keynote,
author = "Mark Harman and Yue Jia and Yuanyuan Zhang",
title = "Achievements, open problems and challenges for search based software testing (keynote Paper)", 
booktitle = "$8^{th}$ {IEEE} International Conference on Software Testing, Verification and Validation ({ICST 2015})", 
address = "Graz, Austria", 
month = "April",
year =  2015
}

@inproceedings{mhpohss:harden,
  title={Harden and Catch for Just-in-Time Assured LLM-Based Software Testing: Open Research Challenges (keynote paper)
},
  author={Harman, Mark and O’Hearn, Peter and Sengupta, Shubho},
  booktitle={2025 {ACM} Conference on Foundations of Software Engineering ({FSE 2025})},
  year={2025},
  organization="{ACM}",
  note={Also available as arXiv preprint arXiv:2504.16472},
}

@inproceedings{foster:mutation,
  title={Mutation-Guided LLM-based Test Generation at Meta},
  author={Foster, Christopher and Gulati, Abhishek and Harman, Mark and Harper, Inna and Mao, Ke and Ritchey, Jillian and Robert, Herv{\'e} and Sengupta, Shubho},
  booktitle={2025 {ACM} Conference on Foundations of Software Engineering ({FSE 2025})},
  year={2025},
  organization="{ACM}",
  note={Also available as arXiv preprint arXiv:2501.12862},
}

@inproceedings{mhetal:TestGen-obs,
    author = "Nadia Alshahwan and Mark Harman and  Alexandru Marginean and Eddy Wang",
    title = "Observation-based unit test generation at Meta",
    booktitle = "Foundations of Software Engineering ({FSE} 2024)",
    year = 2024,

}

@inproceedings{mhetal:fse24-llm,
title = "Automated unit test improvement using {Large Language Models} at {Meta}",
author  = "Nadia Alshahwan and Jubin Chheda and Anastasia Finegenova and Mark Harman and Alexandru Marginean and Shubho Sengupta and Eddy Wang",
booktitle    = "{ACM} International Conference on the Foundations of Software Engineering ({FSE 2024})",
month = "July",
year = 2024,
location = "Porto de Galinhas, Brazil, Brazil",
}

@inproceedings{kinga:enhancing,
  title={Enhancing Testing at {Meta} with Rich-State Simulated Populations},
  author={Alshahwan, Nadia and Blasi, Arianna and Bojarczuk, Kinga and Ciancone, Andrea and Gucevska, Natalija and Harman, Mark and Krolikowski, Michal and Rojas, Rubmary and Martac, Dragos and Schellaert, Simon and others},
  booktitle={Proceedings of the 46th International Conference on Software Engineering: Software Engineering in Practice},
  pages={1--12},
  year={2024}
}

@inproceedings{tuli:simulation,
  author       = {Shreshth Tuli and
                  Kinga Bojarczuk and
                  Natalija Gucevska and
                  Mark Harman and
                  Xiao{-}Yu Wang and
                  Graham Wright},
  title        = {Simulation-Driven Automated End-to-End Test and Oracle Inference},
  booktitle    = {45th {IEEE/ACM} International Conference on Software Engineering:
                  Software Engineering in Practice, SEIP@ICSE 2023, Melbourne, Australia,
                  May 14-20, 2023},
  pages        = {122--133},
  publisher    = {{IEEE}},
  year         = {2023},
}

@inproceedings{kmetal:fausta,
  author    = {Ke Mao and
               Timotej Kapus and
               Lambros Petrou and
               {\'{A}}kos Hajdu and
               Matteo Marescotti and
               Andreas L{\"{o}}scher and
               Mark Harman and
               Dino Distefano},
  title     = {{FAUSTA:} Scaling Dynamic Analysis with Traffic Generation at WhatsApp},
  booktitle = {15th {IEEE} Conference on Software Testing, Verification and Validation,
               {ICST} 2022, Valencia, Spain, April 4-14, 2022},
  pages     = {267--278},
  publisher = {{IEEE}},
  year      = {2022},
}

@inproceedings{jaetal:mia,
author = "John Ahlgren and Maria Eugenia Berezin and Kinga Bojarczuk and Elena Dulskyte and Inna Dvortsova and Johann George and Natalija Gucevska and Mark Harman and Maria Lomeli and Erik Meijer and Silvia Sapora and Justin Spahr-Summers",
title = "Testing Web Enabled Simulation at Scale Using Metamorphic Testing",
booktitle ="International Conference on Software Engineering ({ICSE}) Software Engineering in Practice ({SEIP}) track",
year = 2021,
address = "Virtual"
}

@inproceedings{mike:icse17,
  author    = {Thierry Titcheu Chekam and
               Mike Papadakis and
               Yves Le Traon and
               Mark Harman},
  title     = {An empirical study on mutation, statement and branch coverage fault
               revelation that avoids the unreliable clean program assumption},
  booktitle = {Proceedings of the 39th International Conference on Software Engineering,
               {ICSE} 2017, Buenos Aires, Argentina, May 20-28, 2017},
  pages     = {597--608},
  year      = {2017}
}

@InProceedings{mao:sapienz:16,
author = "Ke Mao and Mark Harman and Yue Jia",
title = "Sapienz: Multi-objective Automated Testing for {Android} Applications",
 booktitle = {International Symposium on Software Testing and Analysis ({ISSTA 2016})},
 year = {2016},
 pages = "94-105"
}

@article{hochberg:sharper,
author = "Yosef Hochberg",
title = "A sharper Bonferroni procedure for multiple tests of significance", 
journal = "Biometrika",
volume = 75,
number = 4,
pages = "800--802",
year = 1988
}

@article{hossain:togll,
  title={{TOGLL}: Correct and strong test oracle generation with {LLMs}},
  author={Hossain, Soneya Binta and Dwyer, Matthew},
  journal={arXiv preprint arXiv:2405.03786},
  year={2024}
}

@article{korel90_dynamic,
	author = {Korel, B.},
	title = {Automated Software Test Data Generation},
	journal = {IEEE Transactions on Software Engineering},
	volume = {16},
	number = {8},
	pages = {870-879},
	year = {1990} 
}

@inproceedings{lakhotia:rule-based,
author = "Arun Lakhotia",
title = "Rule--Based Approach to Computing Module Cohesion",
year = 1993,
booktitle = "$15^{th}$ Conference on Software Engineering
({ICSE-15})",
pages = "34--44"
}

@article{manes:fuzzing,
  author    = {Valentin J. M. Man{\`{e}}s and
               HyungSeok Han and
               Choongwoo Han and
               Sang Kil Cha and
               Manuel Egele and
               Edward J. Schwartz and
               Maverick Woo},
  title     = {The Art, Science, and Engineering of Fuzzing: {A} Survey},
  journal   = {CoRR},
  volume    = {abs/1812.00140},
  year      = {2018},
  archivePrefix = {arXiv},
  eprint    = {1812.00140},
  bibsource = {dblp computer science bibliography, https://dblp.org}
}

@article{mcminn:survey,
author = "Phil {McMinn}",
title = "Search-based Software Test Data Generation: A Survey",
journal = stvr,
pages = "105--156",
Volume =  14, 
Number = 2,
month = jun,
year =  2004
}

@inproceedings{mcminn:past-present-and-future,
  author    = "Phil McMinn",
  title     = "Search-Based Software Testing: Past, Present and Future",
  booktitle = "International Workshop on Search-Based Software Testing ({SBST 2011})", 
  location  = "Berlin, Germany", 
  month     = "21 March", 
  year      = "2011", 
  pages     = "153--163", 
  publisher = "{IEEE}",
  note   = "Keynote paper"
}

@inproceedings{mockus:case,
  title={A case study of open source software development: the Apache server},
  author={Mockus, Audris and Fielding, Roy T and Herbsleb, James},
  booktitle={Proceedings of the 22nd international conference on Software engineering},
  pages={263--272},
  year={2000}
}

@article{molina:test,
  title={Test Oracle Automation in the era of {LLMs}},
  author={Molina, Facundo and Gorla, Alessandra and d’Amorim, Marcelo},
  journal={{ACM} Transactions on Software Engineering and Methodology},
  year={2024},
  publisher={ACM New York, NY}
}

@inproceedings{pacheco:randoop,
  title={Randoop: feedback-directed random testing for Java},
  author={Pacheco, Carlos and Ernst, Michael D},
  booktitle={Companion to the 22nd ACM SIGPLAN conference on Object-oriented programming systems and applications companion},
  pages={815--816},
  year={2007}
}

@article{perneger1998whats,
  title={What's wrong with {B}onferroni adjustments},
  author={Perneger, Thomas V},
  journal={British Medical Journal},
  volume={316},
  number={7139},
  pages={1236--1238},
  year={1998},
  publisher={British Medical Journal Publishing Group}
}

@book{pearl2009causality,
  title={Causality},
  author={Pearl, Judea},
  year={2009},
  publisher={Cambridge university press}
}

@article{pizzorno:coverup,
  title={Coverup: Coverage-guided {LLM}-based test generation},
  author={Pizzorno, Juan Altmayer and Berger, Emery D},
  journal={arXiv preprint arXiv:2403.16218},
  year={2024}
}

@article{rothman1990no,
  title={No adjustments are needed for multiple comparisons},
  author={Rothman, Kenneth J},
  journal={Epidemiology},
  volume={1},
  number={1},
  pages={43--46},
  year={1990},
  publisher={LWW}
}

@inproceedings{sadowski:modern,
  title={Modern code review: a case study at {Google}},
  author={Sadowski, Caitlin and S{\"o}derberg, Emma and Church, Luke and Sipko, Michal and Bacchelli, Alberto},
  booktitle={Proceedings of the 40th International Conference on Software Engineering: Software Engineering in Practice},
  pages={181--190},
  year={2018}
}

@InProceedings{sen:cute,
  title =	"{CUTE}: a concolic unit testing engine for {C}",
  author =	"Koushik Sen and Darko Marinov and Gul Agha",
  booktitle =	"$10^{th}$ European Software Engineering
		 Conference and 13th {ACM} 
		 International Symposium on Foundations of Software
		 Engineering ({ESEC/FSE} '05)",
  publisher =	"ACM",
  year = 	"2005",
  editor =	"Michel Wermelinger and Harald Gall",
  ISBN = 	"1-59593-014-0",
  pages =	"263--272"}

@book{shepperd:foundations,
author = "Martin J. Shepperd",
title = "Foundations of software measurement",
publisher = "Prentice Hall",
year = 1995
}

@article{simpson1951interpretation,
  title={The interpretation of interaction in contingency tables},
  author={Simpson, Edward H},
  journal={Journal of the Royal Statistical Society: Series {B} (Methodological)},
  volume={13},
  number={2},
  pages={238--241},
  year={1951},
  publisher={Wiley Online Library}
}

@Article{weyuker:untestable,
  author =       "Elaine J. Weyuker",
  title =        "On Testing Non-Testable Programs",
  journal =      "The Computer Journal",
  volume =       "25",
  number =       "4",
  pages =        "465--470",
  month =        nov,
  year =         "1982",
  ISSN =         "0010-4620"
}

@misc{otten:prompting,
      title={Prompting in Practice: Investigating Software Developers' Use of Generative {AI} Tools}, 
      author={Daniel Otten and Trevor Stalnaker and Nathan Wintersgill and Oscar Chaparro and Denys Poshyvanyk},
      year={2025},
      eprint={2510.06000},
      archivePrefix={arXiv},
      primaryClass={cs.SE},
      url={https://arxiv.org/abs/2510.06000}, 
}

@article{xiao:generative,
   title={Generative {AI} for Pull Request Descriptions: Adoption, Impact, and Developer Interventions},
   volume={1},
   ISSN={2994-970X},
   number={FSE},
   journal={Proceedings of the {ACM} on Software Engineering},
   publisher={Association for Computing Machinery ({ACM})},
   author={Xiao, Tao and Hata, Hideaki and Treude, Christoph and Matsumoto, Kenichi},
   year={2024},
   month=jul, pages={1043–1065} }

@book{zeller:programs,
  title={Why programs fail: a guide to systematic debugging},
  author={Zeller, Andreas},
  year={2009},
  publisher={Morgan Kaufmann}
}

@InProceedings{zhang:empirically,
title =	"Empirically revisiting the test independence assumption",
author =	"Sai Zhang and Darioush Jalali and Jochen Wuttke and Kivan{\c c} Muslu and Wing Lam and Michael D. Ernst and David Notkin",
booktitle =	"International Symposium on Software Testing and Analysis ({ISSTA 2014})", 
address = "San Jose, CA, USA",
month = "July 21 - 26",
publisher =	"ACM",
year = 	"2014",
editor =	"Corina S. Pasareanu and Darko Marinov",
ISBN = 	"978-1-4503-2645-2",
pages =	"385--396"
}

@inproceedings{movefast,
  author    = {C. Calcagno and
               D. Distefano and
               J. Dubreil and
               D. Gabi and
               P. Hooimeijer and
               M. Luca and
               P. W. O'Hearn and
               I. Papakonstantinou and
               J. Purbrick and
               D. Rodriguez},
  title     = {Moving Fast with Software Verification},
  booktitle = {{NASA} Formal Methods - 7th International Symposium},
  pages     = {3--11},
  year      = {2015}
}

@article{zheng2023judging,
  title={Judging {LLM}-as-a-judge with {MT}-bench and chatbot arena},
  author={Zheng, Lianmin and Chiang, Wei-Lin and Sheng, Ying and Zhuang, Siyuan and Wu, Zhanghao and Zhuang, Yonghao and Lin, Zi and Li, Zhuohan and Li, Dacheng and Xing, Eric and others},
  journal={Advances in Neural Information Processing Systems ({NeurIPS 2023})},
  volume={36},
  pages={46595--46623},
  year={2023}
}

@misc{schafer_adaptive_2023,
    title = {Adaptive {Test} {Generation} {Using} a {Large} {Language} {Model}},
    language = {en},
    publisher = {arXiv},
    author = {Schäfer, Max and Nadi, Sarah and Eghbali, Aryaz and Tip, Frank},
    month = feb,
    year = {2023},
    note = {arXiv:2302.06527  },
}

@misc{white_chatgpt_2023,
    title = {{ChatGPT} {Prompt} {Patterns} for {Improving} {Code} {Quality}, {Refactoring}, {Requirements} {Elicitation}, and {Software} {Design}},
    language = {en},
    publisher = {arXiv},
    author = {White, Jules and Hays, Sam and Fu, Quchen and Spencer-Smith, Jesse and Schmidt, Douglas C.},
    month = mar,
    year = {2023},
    note = {arXiv:2303.07839  },
}

@article{shin2023prompt,
  title={Prompt Engineering or Fine Tuning: An Empirical Assessment of Large Language Models in Automated Software Engineering Tasks},
  author={Shin, Jiho and Tang, Clark and Mohati, Tahmineh and Nayebi, Maleknaz and Wang, Song and Hemmati, Hadi},
  journal={arXiv preprint arXiv:2310.10508},
  year={2023}
}

@misc{wang:llm-test-survey,
      title={Software Testing with Large Language Model: Survey, Landscape, and Vision}, 
      author={Junjie Wang and Yuchao Huang and Chunyang Chen and Zhe Liu and Song Wang and Qing Wang},
      year={2023},
      note={arXiv:2307.07221},
      archivePrefix={arXiv},
      primaryClass={cs.SE}
}

@inproceedings{ibrahimzada2022perfect,
  title={Perfect is the enemy of test oracle},
  author={Ibrahimzada, Ali Reza and Varli, Yigit and Tekinoglu, Dilara and Jabbarvand, Reyhaneh},
  booktitle={Proceedings of the 30th {ACM} Joint {European} Software Engineering Conference and Symposium on the Foundations of Software Engineering},
  pages={70--81},
  year={2022}
}

@article{ryan2024code,
  title={Code-aware prompting: A study of coverage-guided test generation in regression setting using {LLM}},
  author={Ryan, Gabriel and Jain, Siddhartha and Shang, Mingyue and Wang, Shiqi and Ma, Xiaofei and Ramanathan, Murali Krishna and Ray, Baishakhi},
  journal={Proceedings of the ACM on Conference on Foundations of  Software Engineering},
  volume={1},
  number={FSE},
  pages={951--971},
  year={2024},
  publisher={ACM New York, NY, USA}
}

@inproceedings{chen2024chatunitest,
  title={Chatunitest: A framework for {LLM}-based test generation},
  author={Chen, Yinghao and Hu, Zehao and Zhi, Chen and Han, Junxiao and Deng, Shuiguang and Yin, Jianwei},
  booktitle={Companion Proceedings of the 32nd ACM International Conference on the Foundations of Software Engineering},
  pages={572--576},
  year={2024}
}

@article{schafer2023empirical,
  title={An empirical evaluation of using large language models for automated unit test generation},
  author={Sch{\"a}fer, Max and Nadi, Sarah and Eghbali, Aryaz and Tip, Frank},
  journal={{IEEE} Transactions on Software Engineering},
  year={2023},
  publisher={IEEE}
}

@article{liu2024llm,
      title={{LLM}-Powered Test Case Generation for Detecting Bugs in Plausible Programs}, 
      author={Kaibo Liu and Zhenpeng Chen and Yiyang Liu and Jie M. Zhang and Mark Harman and Yudong Han and Yun Ma and Yihong Dong and Ge Li and Gang Huang},
      year={2025},
      eprint={2404.10304},
      archivePrefix={arXiv},
      primaryClass={cs.SE},
      url={https://arxiv.org/abs/2404.10304}, 
}

\newpage
\appendix

%%% START OF AUTO GENERATED TEXT ... DO NOT MANUALLY EDIT ...
%%% START OF AUTO GENERATED TEXT ... DO NOT MANUALLY EDIT ...
%%% START OF AUTO GENERATED TEXT ... DO NOT MANUALLY EDIT ...
%%% START OF AUTO GENERATED TEXT ... DO NOT MANUALLY EDIT ...
%%% START OF AUTO GENERATED TEXT ... DO NOT MANUALLY EDIT ...
%%% START OF AUTO GENERATED TEXT ... DO NOT MANUALLY EDIT ...
%%% START OF AUTO GENERATED TEXT ... DO NOT MANUALLY EDIT ...
%%% START OF AUTO GENERATED TEXT ... DO NOT MANUALLY EDIT ...
%%% START OF AUTO GENERATED TEXT ... DO NOT MANUALLY EDIT ...
%%% START OF AUTO GENERATED TEXT ... DO NOT MANUALLY EDIT ...
%%% START OF AUTO GENERATED TEXT ... DO NOT MANUALLY EDIT ...

%% ============================================================================
%% LaTeX output generated by ACH Analysis
%% ----------------------------------------------------------------------------
%% Command:  /data/sandcastle/boxes/fbsource/buck-out/v2/gen/fbcode/0aecfa677e269d14/automate_ig/simplification/__llamoptimiser__/llamoptimiser.par --analysis --latex --include-assessment-pvalue-matrices --fisher-randomization-iterations=100000 --apply-llm-test-quality-filter
%% Created:  2026-01-08 01:29:04
%% User:     markharman
%% ============================================================================

% Fisher's Exact Test Randomization Sense Check
% ------------------------------------------------------------
% Fisher's Exact Test Randomization Sense Check
% ------------------------------------------------------------

\begin{table} 
\centering
\begin{tabular}{lccc}
\toprule
 & \textbf{RubFake} & \textbf{TP Prob} & \textbf{Bucket Med} \\
\midrule
\multicolumn{4}{l}{\textit{True Positive (score $> 0$)}} \\
Significant & 4271/100000 & 4359/100000 & 4379/100000 \\
Small & 2977 & 3839 & 3544 \\
Medium & 0 & 0 & 0 \\
Large & 0 & 0 & 0 \\
Prob. $\geq$ L & 0.00000 & 0.00000 & 0.00000 \\
Prob. $\geq$ M & 0.00000 & 0.00000 & 0.00000 \\
Prob. $\geq$ S & 0.02977 & 0.03839 & 0.03544 \\
Prob. $\geq$ N & 0.04271 & 0.04359 & 0.04379 \\
\midrule
\multicolumn{4}{l}{\textit{False Positive (score $< 0$)}} \\
Significant & 4243/100000 & 4262/100000 & 4629/100000 \\
Small & 3332 & 3767 & 3904 \\
Medium & 0 & 0 & 0 \\
Large & 0 & 0 & 0 \\
Prob. $\geq$ L & 0.00000 & 0.00000 & 0.00000 \\
Prob. $\geq$ M & 0.00000 & 0.00000 & 0.00000 \\
Prob. $\geq$ S & 0.03332 & 0.03767 & 0.03904 \\
Prob. $\geq$ N & 0.04243 & 0.04262 & 0.04629 \\
\bottomrule
\end{tabular}
\caption{Fisher's Exact Test randomization sense check for Good and Bad diff status groups. Each cell shows the number of significant results ($p < 0.05$) from 100000 tests on randomly shuffled data. Each shuffled iteration simulates arbitrary good and bad sets by ignoring labels, with set sizes ranging from 205 to 232 (mean: 218.5). Small, Medium, and Large rows show counts of significant results at each effect size level. Cumulative probability rows (Prob. $\geq$ X) estimate the probability of achieving an effect size at or above level X by chance (e.g., Prob. $\geq$ M = (L + M) / 100000). Expected false positives by chance: 5000 (5\%).}
\label{tab:fisher-randomization-sanity-check}
\end{table}

% Small Sample Fisher's Exact Test Randomization Sense Check
% ------------------------------------------------------------
% Small Sample Fisher's Exact Test Randomization Sense Check
% ------------------------------------------------------------

\begin{table} 
\centering
\begin{tabular}{lccc}
\toprule
 & \textbf{RubFake} & \textbf{TP Prob} & \textbf{Bucket Med} \\
\midrule
\multicolumn{4}{l}{\textit{True Positive (score $> 0$)}} \\
Significant & 2274/100000 & 3216/100000 & 2616/100000 \\
Small & 48 & 122 & 62 \\
Medium & 1185 & 1842 & 1362 \\
Large & 1041 & 1252 & 1192 \\
Prob. $\geq$ L & 0.01041 & 0.01252 & 0.01192 \\
Prob. $\geq$ M & 0.02226 & 0.03094 & 0.02554 \\
Prob. $\geq$ S & 0.02274 & 0.03216 & 0.02616 \\
\midrule
\multicolumn{4}{l}{\textit{False Positive (score $< 0$)}} \\
Significant & 3349/100000 & 3350/100000 & 3320/100000 \\
Small & 159 & 150 & 108 \\
Medium & 1917 & 1860 & 1893 \\
Large & 1273 & 1340 & 1319 \\
Prob. $\geq$ L & 0.01273 & 0.01340 & 0.01319 \\
Prob. $\geq$ M & 0.03190 & 0.03200 & 0.03212 \\
Prob. $\geq$ S & 0.03349 & 0.03350 & 0.03320 \\
\bottomrule
\end{tabular}
\caption{Small sample Fisher's Exact Test randomization sense check. Each cell shows the number of significant results ($p < 0.05$) from 100000 tests on randomly shuffled data. Each shuffled iteration simulates arbitrary comparisons using sample sizes ranging from 6 to 48 per group, representative of the smallest status pairs that showed significant results with effect size $\geq$ medium in Tables~\ref{tab:tp-rate-pvalue-matrices-by-approach} and~\ref{tab:fp-rate-pvalue-matrices-by-approach}. Small, Medium, and Large rows show counts of significant results at each effect size level. Cumulative probability rows (Prob. $\geq$ X) estimate the probability of achieving an effect size at or above level X by chance. Expected false positives by chance: 5000 (5\%).}
\label{tab:fisher-randomization-small-sample}
\end{table}

% Inter-Rater Agreement and Rank Correlation Analysis

%------------------------------------------------------------
% Inter-Rater Agreement and Rank Correlation Analysis
%------------------------------------------------------------

% Assessment Metrics (Mann-Whitney U)

%------------------------------------------------------------
% Grouped P-value Matrices for Assessment Metrics
% NOTE: Requires \usepackage{graphicx} for \rotatebox and \usepackage{subcaption} for subtables
%------------------------------------------------------------

\begin{table}
\centering
\caption{P-value matrices for assessment metrics comparing diff status pairs (upper triangle). Bold values are significant at $\alpha=0.05$. $\leftarrow P$ = row has higher scores; $\uparrow P$ = column has higher scores. Effect size: N=negligible, S=small, M=medium, L=large. Status abbreviations: CLS=CLOSED, ACC=ACCEPTED, ABD=ABANDONED, CHP=CHANGES\_PLANNED, NRS=NEEDS\_REVISION, REV=REVERTED.}
\label{tab:pvalue-assessment-metrics}

\begin{subtable}{0.45\textwidth}
\centering
\resizebox{\linewidth}{!}{%
\begin{tabular}{lcccccc}
\hline
 & \rotatebox{45}{ACC} & \rotatebox{45}{ABD} & \rotatebox{45}{CHP} & \rotatebox{45}{NRS} & \rotatebox{45}{REV} & \rotatebox{45}{Bad}\hspace{-3pt}\rotatebox{45}{(n=107)}\hspace{-3pt}\rotatebox{45}{(0.0\%)} \\
\hline
CLS (n=91) & 0.197 & 0.260 & 0.429 & 0.895 & N/A &  \\
ACC (n=62) &  & 0.919 & 0.975 & 0.381 & N/A &  \\
ABD (n=77) &  &  & 0.935 & 0.466 & N/A &  \\
CHP (n=12) &  &  &  & 0.568 & N/A &  \\
NRS (n=17) &  &  &  &  & N/A &  \\
REV (n=1) &  &  &  &  &  &  \\
\hline
G (n=153) &  &  &  &  &  & 0.633 \\
\hline
\multicolumn{7}{l}{Overall total n = 260 diffs} \\
\hline
\end{tabular}
}
\vspace{0.3em}\centering\small{(rubfake\_overall\_likelihood\_score. Mann-Whitney U test; effect size: Cliff's $\delta$.)}
\end{subtable}

\vspace{0.5em}
\begin{subtable}{0.45\textwidth}
\centering
\resizebox{\linewidth}{!}{%
\begin{tabular}{lcccccc}
\hline
 & \rotatebox{45}{ACC} & \rotatebox{45}{ABD} & \rotatebox{45}{CHP} & \rotatebox{45}{NRS} & \rotatebox{45}{REV} & \rotatebox{45}{Bad}\hspace{-3pt}\rotatebox{45}{(n=107)}\hspace{-3pt}\rotatebox{45}{(0.0\%)} \\
\hline
CLS (n=91) & 0.527 & $\uparrow$\textbf{0.002} (S) & 0.557 & 0.385 & N/A &  \\
ACC (n=62) &  & $\uparrow$\textbf{0.038} (S) & 0.778 & 0.763 & N/A &  \\
ABD (n=77) &  &  & 0.235 & 0.302 & N/A &  \\
CHP (n=12) &  &  &  & 0.710 & N/A &  \\
NRS (n=17) &  &  &  &  & N/A &  \\
REV (n=1) &  &  &  &  &  &  \\
\hline
G (n=153) &  &  &  &  &  & $\uparrow$\textbf{0.005} (S) \\
\hline
\multicolumn{7}{l}{Overall total n = 260 diffs} \\
\hline
\end{tabular}
}
\vspace{0.3em}\centering\small{(true\_positive\_probability. Mann-Whitney U test; effect size: Cliff's $\delta$.)}
\end{subtable}

\vspace{0.5em}
\begin{subtable}{0.45\textwidth}
\centering
\resizebox{\linewidth}{!}{%
\begin{tabular}{lcccccc}
\hline
 & \rotatebox{45}{ACC} & \rotatebox{45}{ABD} & \rotatebox{45}{CHP} & \rotatebox{45}{NRS} & \rotatebox{45}{REV} & \rotatebox{45}{Bad}\hspace{-3pt}\rotatebox{45}{(n=107)}\hspace{-3pt}\rotatebox{45}{(0.0\%)} \\
\hline
CLS (n=91) & 0.922 & 0.071 & 0.225 & 0.120 & N/A &  \\
ACC (n=62) &  & 0.076 & 0.214 & 0.117 & N/A &  \\
ABD (n=77) &  &  & 0.889 & 0.737 & N/A &  \\
CHP (n=12) &  &  &  & 0.917 & N/A &  \\
NRS (n=17) &  &  &  &  & N/A &  \\
REV (n=1) &  &  &  &  &  &  \\
\hline
G (n=153) &  &  &  &  &  & $\uparrow$\textbf{0.014} (S) \\
\hline
\multicolumn{7}{l}{Overall total n = 260 diffs} \\
\hline
\end{tabular}
}
\vspace{0.3em}\centering\small{(true\_positive\_bucket\_median\_score. Mann-Whitney U test; effect size: Cliff's $\delta$.)}
\end{subtable}

\vspace{0.5em}
\end{table}

% Catch Rate Approaches (Fisher's Exact Test)

%------------------------------------------------------------
% Grouped P-value Matrices for Catch Rate Approaches
% NOTE: Requires \usepackage{graphicx} for \rotatebox and \usepackage{subcaption} for subtables
%------------------------------------------------------------

\begin{table}
\centering
\caption{Catch rate P-value matrices comparing diff status pairs (upper triangle). P-values computed using Fisher's Exact Test; effect sizes using Cohen's $h$ \cite{cohen:statistical-power}. Bold values are significant at $\alpha=0.05$. $\leftarrow P$ = row catch rate higher; $\uparrow P$ = column catch rate higher. n = total samples, (\%) = catch rate. Effect size (Cohen's $h$): N=negligible ($|h|<0.2$), S=small ($0.2 \leq |h|<0.5$), M=medium ($0.5 \leq |h|<0.8$), L=large ($|h| \geq 0.8$). Status abbreviations: CLS=CLOSED, ACC=ACCEPTED, ABD=ABANDONED, CHP=CHANGES\_PLANNED, NRS=NEEDS\_REVISION, REV=REVERTED.}
\label{tab:catch-rate-approaches}

\begin{subtable}{0.45\textwidth}
\centering
\resizebox{\linewidth}{!}{%
\begin{tabular}{lcccccc}
\hline
 & \rotatebox{45}{ACC} & \rotatebox{45}{ABD} & \rotatebox{45}{CHP} & \rotatebox{45}{NRS} & \rotatebox{45}{REV} & \rotatebox{45}{Bad}\hspace{-3pt}\rotatebox{45}{(n=398)} \\
\hline
CLS n=249 (18.5\%) & 0.377 & $\leftarrow$\textbf{0.003} (S) & 0.263 & 0.552 & 0.463 &  \\
ACC n=198 (15.2\%) &  & 0.065 & 0.467 & 0.282 & 0.397 &  \\
ABD n=305 (9.5\%) &  &  & 1.000 & $\uparrow$\textbf{0.014} (S) & 0.265 &  \\
CHP n=41 (9.8\%) &  &  &  & 0.156 & 0.310 &  \\
NRS n=49 (22.4\%) &  &  &  &  & 0.553 &  \\
REV n=3 (33.3\%) &  &  &  &  &  &  \\
\hline
Good n=447 (17.0\%) &  &  &  &  &  & $\leftarrow$\textbf{0.018} (N) \\
\hline
\multicolumn{7}{l}{Overall total: n=845 (14.3\% of which are catches, which is 121 catches in total)} \\
\hline
\end{tabular}
}
\vspace{0.3em}\centering\small{(intent-aware workflow)}
\end{subtable}

\vspace{0.5em}
\begin{subtable}{0.45\textwidth}
\centering
\resizebox{\linewidth}{!}{%
\begin{tabular}{lcccccc}
\hline
 & \rotatebox{45}{ACC} & \rotatebox{45}{ABD} & \rotatebox{45}{CHP} & \rotatebox{45}{NRS} & \rotatebox{45}{REV} & \rotatebox{45}{Bad}\hspace{-3pt}\rotatebox{45}{(n=543)} \\
\hline
CLS n=387 (11.6\%) & 0.715 & 0.740 & 0.371 & 1.000 & 1.000 &  \\
ACC n=305 (10.5\%) &  & 0.405 & 0.565 & 1.000 & 1.000 &  \\
ABD n=381 (12.6\%) &  &  & 0.224 & 0.830 & 1.000 &  \\
CHP n=103 (7.8\%) &  &  &  & 0.569 & 1.000 &  \\
NRS n=57 (10.5\%) &  &  &  &  & 1.000 &  \\
REV n=2 (0.0\%) &  &  &  &  &  &  \\
\hline
Good n=692 (11.1\%) &  &  &  &  &  & 0.928 \\
\hline
\multicolumn{7}{l}{Overall total: n=1235 (11.3\% of which are catches, which is 139 catches in total)} \\
\hline
\end{tabular}
}
\vspace{0.3em}\centering\small{(dodgy diff workflow)}
\end{subtable}

\vspace{0.5em}
\begin{subtable}{0.45\textwidth}
\centering
\resizebox{\linewidth}{!}{%
\begin{tabular}{lcccccc}
\hline
 & \rotatebox{45}{ACC} & \rotatebox{45}{ABD} & \rotatebox{45}{CHP} & \rotatebox{45}{NRS} & \rotatebox{45}{REV} & \rotatebox{45}{Bad}\hspace{-3pt}\rotatebox{45}{(n=941)} \\
\hline
CLS n=636 (14.3\%) & 0.337 & 0.099 & 0.057 & 0.656 & 0.540 &  \\
ACC n=503 (12.3\%) &  & 0.584 & 0.234 & 0.339 & 0.486 &  \\
ABD n=686 (11.2\%) &  &  & 0.375 & 0.150 & 0.452 &  \\
CHP n=144 (8.3\%) &  &  &  & 0.073 & 0.371 &  \\
NRS n=106 (16.0\%) &  &  &  &  & 1.000 &  \\
REV n=5 (20.0\%) &  &  &  &  &  &  \\
\hline
Good n=1139 (13.4\%) &  &  &  &  &  & 0.162 \\
\hline
\multicolumn{7}{l}{Overall total: n=2080 (12.5\% of which are catches, which is 260 catches in total)} \\
\hline
\end{tabular}
}
\vspace{0.3em}\centering\small{(All Approaches Combined)}
\end{subtable}

\vspace{0.5em}
\end{table}

% TP Rate by Status (Fisher's Exact Test)

%------------------------------------------------------------
% Grouped P-value Matrices for TP Rate by Diff Status and Approach
% NOTE: Requires \usepackage{graphicx} for \rotatebox and \usepackage{subcaption} for subtables
%------------------------------------------------------------

\begin{table*} 
\centering

% --- All Approaches ---
\textbf{All Approaches:}
\\[0.3em]
\begin{subtable}{0.32\textwidth}
\centering
\resizebox{\linewidth}{!}{%
\begin{tabular}{lccccc}
\hline
 & \rotatebox{45}{ACC} & \rotatebox{45}{ABD} & \rotatebox{45}{CHP} & \rotatebox{45}{NRS} & \rotatebox{45}{REV} \\
\hline
CLS n=91 (11.0\%) & 0.243 & 0.368 & 1.000 & 0.427 & 1.000 \\
ACC n=62 (4.8\%) &  & $\uparrow$\textbf{0.033} (S) & 0.515 & 0.110 & 1.000 \\
ABD n=77 (16.9\%) &  &  & 0.682 & 1.000 & 1.000 \\
CHP n=12 (8.3\%) &  &  &  & 0.622 & 1.000 \\
NRS n=17 (17.6\%) &  &  &  &  & 1.000 \\
REV n=1 (0.0\%) &  &  &  &  &  \\
\hline
\end{tabular}
}
\subcaption{rubfake}
\end{subtable}
\hfill
\begin{subtable}{0.32\textwidth}
\centering
\resizebox{\linewidth}{!}{%
\begin{tabular}{lccccc}
\hline
 & \rotatebox{45}{ACC} & \rotatebox{45}{ABD} & \rotatebox{45}{CHP} & \rotatebox{45}{NRS} & \rotatebox{45}{REV} \\
\hline
CLS n=90 (37.8\%) & 0.178 & $\uparrow$\textbf{0.003} (S) & 1.000 & 0.411 & 1.000 \\
ACC n=60 (50.0\%) &  & 0.226 & 0.754 & 1.000 & 1.000 \\
ABD n=77 (61.0\%) &  &  & 0.225 & 0.420 & 0.397 \\
CHP n=12 (41.7\%) &  &  &  & 0.718 & 1.000 \\
NRS n=16 (50.0\%) &  &  &  &  & 1.000 \\
REV n=1 (0.0\%) &  &  &  &  &  \\
\hline
\end{tabular}
}
\subcaption{probability}
\end{subtable}
\hfill
\begin{subtable}{0.32\textwidth}
\centering
\resizebox{\linewidth}{!}{%
\begin{tabular}{lccccc}
\hline
 & \rotatebox{45}{ACC} & \rotatebox{45}{ABD} & \rotatebox{45}{CHP} & \rotatebox{45}{NRS} & \rotatebox{45}{REV} \\
\hline
CLS n=90 (10.0\%) & 0.564 & $\uparrow$\textbf{0.022} (S) & 0.615 & 0.386 & 1.000 \\
ACC n=60 (6.7\%) &  & $\uparrow$\textbf{0.009} (S) & 0.260 & 0.157 & 1.000 \\
ABD n=77 (23.4\%) &  &  & 1.000 & 1.000 & 1.000 \\
CHP n=12 (16.7\%) &  &  &  & 1.000 & 1.000 \\
NRS n=16 (18.8\%) &  &  &  &  & 1.000 \\
REV n=1 (0.0\%) &  &  &  &  &  \\
\hline
\end{tabular}
}
\subcaption{bucket\_median\_score}
\end{subtable}
\\[1em]

% --- make_and_kill_mutants_from_diff_intent_and_risks ---
\textbf{intent-aware workflow:}
\\[0.3em]
\begin{subtable}{0.32\textwidth}
\centering
\resizebox{\linewidth}{!}{%
\begin{tabular}{lccccc}
\hline
 & \rotatebox{45}{ACC} & \rotatebox{45}{ABD} & \rotatebox{45}{CHP} & \rotatebox{45}{NRS} & \rotatebox{45}{REV} \\
\hline
CLS n=46 (10.9\%) & 0.697 & 0.700 & 0.411 & 0.571 & 1.000 \\
ACC n=30 (6.7\%) &  & 1.000 & 0.322 & 1.000 & 1.000 \\
ABD n=29 (6.9\%) &  &  & 0.330 & 1.000 & 1.000 \\
CHP n=4 (25.0\%) &  &  &  & 0.267 & 1.000 \\
NRS n=11 (0.0\%) &  &  &  &  & 1.000 \\
REV n=1 (0.0\%) &  &  &  &  &  \\
\hline
\end{tabular}
}
\subcaption{rubfake}
\end{subtable}
\hfill
\begin{subtable}{0.32\textwidth}
\centering
\resizebox{\linewidth}{!}{%
\begin{tabular}{lccccc}
\hline
 & \rotatebox{45}{ACC} & \rotatebox{45}{ABD} & \rotatebox{45}{CHP} & \rotatebox{45}{NRS} & \rotatebox{45}{REV} \\
\hline
CLS n=45 (31.1\%) & $\uparrow$\textbf{0.016} (M) & $\uparrow$\textbf{0.005} (M) & 0.588 & 0.706 & 1.000 \\
ACC n=28 (60.7\%) &  & 0.787 & 1.000 & 0.062 & 0.414 \\
ABD n=29 (65.5\%) &  &  & 0.610 & $\leftarrow$\textbf{0.025} (L) & 0.367 \\
CHP n=4 (50.0\%) &  &  &  & 0.520 & 1.000 \\
NRS n=10 (20.0\%) &  &  &  &  & 1.000 \\
REV n=1 (0.0\%) &  &  &  &  &  \\
\hline
\end{tabular}
}
\subcaption{probability}
\end{subtable}
\hfill
\begin{subtable}{0.32\textwidth}
\centering
\resizebox{\linewidth}{!}{%
\begin{tabular}{lccccc}
\hline
 & \rotatebox{45}{ACC} & \rotatebox{45}{ABD} & \rotatebox{45}{CHP} & \rotatebox{45}{NRS} & \rotatebox{45}{REV} \\
\hline
CLS n=45 (8.9\%) & 0.473 & 0.051 & 1.000 & 1.000 & 1.000 \\
ACC n=28 (14.3\%) &  & 0.331 & 1.000 & 1.000 & 1.000 \\
ABD n=29 (27.6\%) &  &  & 0.550 & 0.400 & 1.000 \\
CHP n=4 (0.0\%) &  &  &  & 1.000 & 1.000 \\
NRS n=10 (10.0\%) &  &  &  &  & 1.000 \\
REV n=1 (0.0\%) &  &  &  &  &  \\
\hline
\end{tabular}
}
\subcaption{bucket\_median\_score}
\end{subtable}
\\[1em]

% --- make_tests_from_parent_and_dodgy_diff ---
\textbf{dodgy diff workflow:}
\\[0.3em]
\begin{subtable}{0.32\textwidth}
\centering
\resizebox{\linewidth}{!}{%
\begin{tabular}{lcccc}
\hline
 & \rotatebox{45}{ACC} & \rotatebox{45}{ABD} & \rotatebox{45}{CHP} & \rotatebox{45}{NRS} \\
\hline
CLS n=45 (11.1\%) & 0.391 & 0.173 & 1.000 & $\uparrow$\textbf{0.042} (L) \\
ACC n=32 (3.1\%) &  & $\uparrow$\textbf{0.023} (M) & 1.000 & $\uparrow$\textbf{0.009} (L) \\
ABD n=48 (22.9\%) &  &  & 0.333 & 0.173 \\
CHP n=8 (0.0\%) &  &  &  & 0.055 \\
NRS n=6 (50.0\%) &  &  &  &  \\
\hline
\end{tabular}
}
\subcaption{rubfake}
\end{subtable}
\hfill
\begin{subtable}{0.32\textwidth}
\centering
\resizebox{\linewidth}{!}{%
\begin{tabular}{lcccc}
\hline
 & \rotatebox{45}{ACC} & \rotatebox{45}{ABD} & \rotatebox{45}{CHP} & \rotatebox{45}{NRS} \\
\hline
CLS n=45 (44.4\%) & 0.817 & 0.216 & 1.000 & $\uparrow$\textbf{0.023} (L) \\
ACC n=32 (40.6\%) &  & 0.171 & 1.000 & $\uparrow$\textbf{0.020} (L) \\
ABD n=48 (58.3\%) &  &  & 0.445 & 0.074 \\
CHP n=8 (37.5\%) &  &  &  & $\uparrow$\textbf{0.031} (L) \\
NRS n=6 (100.0\%) &  &  &  &  \\
\hline
\end{tabular}
}
\subcaption{probability}
\end{subtable}
\hfill
\begin{subtable}{0.32\textwidth}
\centering
\resizebox{\linewidth}{!}{%
\begin{tabular}{lcccc}
\hline
 & \rotatebox{45}{ACC} & \rotatebox{45}{ABD} & \rotatebox{45}{CHP} & \rotatebox{45}{NRS} \\
\hline
CLS n=45 (11.1\%) & 0.072 & 0.264 & 0.283 & 0.186 \\
ACC n=32 (0.0\%) &  & $\uparrow$\textbf{0.005} (L) & $\uparrow$\textbf{0.036} (L) & $\uparrow$\textbf{0.021} (L) \\
ABD n=48 (20.8\%) &  &  & 1.000 & 0.605 \\
CHP n=8 (25.0\%) &  &  &  & 1.000 \\
NRS n=6 (33.3\%) &  &  &  &  \\
\hline
\end{tabular}
}
\subcaption{bucket\_median\_score}
\end{subtable}

\caption{P-value matrices for true positive rates (score $> 0$) comparing diff status pairs (upper triangle), grouped by catch generation approach. P-values computed using Fisher's Exact Test; effect sizes using Cohen's $h$ \cite{cohen:statistical-power}. Bold values are significant at $\alpha=0.05$. $\leftarrow P$ = row has higher TP rate; $\uparrow P$ = column has higher TP rate. Effect size (Cohen's $h$): N=negligible ($|h|<0.2$), S=small ($0.2 \leq |h|<0.5$), M=medium ($0.5 \leq |h|<0.8$), L=large ($|h| \geq 0.8$). Status abbreviations: CLS=CLOSED, ACC=ACCEPTED, ABD=ABANDONED, CHP=CHANGES\_PLANNED, NRS=NEEDS\_REVISION, REV=REVERTED.}
\label{tab:tp-rate-pvalue-matrices-by-approach}
\end{table*}

% FP Rate by Status (Fisher's Exact Test)

%------------------------------------------------------------
% Grouped P-value Matrices for FP Rate by Diff Status and Approach
% NOTE: Requires \usepackage{graphicx} for \rotatebox and \usepackage{subcaption} for subtables
%------------------------------------------------------------

\begin{table*} 
\centering

% --- All Approaches ---
\textbf{All Approaches:}
\\[0.3em]
\begin{subtable}{0.32\textwidth}
\centering
\resizebox{\linewidth}{!}{%
\begin{tabular}{lccccc}
\hline
 & \rotatebox{45}{ACC} & \rotatebox{45}{ABD} & \rotatebox{45}{CHP} & \rotatebox{45}{NRS} & \rotatebox{45}{REV} \\
\hline
CLS n=91 (51.6\%) & 0.622 & 0.878 & 0.373 & 0.795 & 0.489 \\
ACC n=62 (56.5\%) &  & 0.734 & 0.750 & 0.586 & 0.444 \\
ABD n=77 (53.2\%) &  &  & 0.536 & 0.790 & 0.474 \\
CHP n=12 (66.7\%) &  &  &  & 0.451 & 0.385 \\
NRS n=17 (47.1\%) &  &  &  &  & 1.000 \\
REV n=1 (0.0\%) &  &  &  &  &  \\
\hline
\end{tabular}
}
\subcaption{rubfake}
\end{subtable}
\hfill
\begin{subtable}{0.32\textwidth}
\centering
\resizebox{\linewidth}{!}{%
\begin{tabular}{lccccc}
\hline
 & \rotatebox{45}{ACC} & \rotatebox{45}{ABD} & \rotatebox{45}{CHP} & \rotatebox{45}{NRS} & \rotatebox{45}{REV} \\
\hline
CLS n=90 (62.2\%) & 0.178 & $\leftarrow$\textbf{0.003} (S) & 1.000 & 0.411 & 1.000 \\
ACC n=60 (50.0\%) &  & 0.226 & 0.754 & 1.000 & 1.000 \\
ABD n=77 (39.0\%) &  &  & 0.225 & 0.420 & 0.397 \\
CHP n=12 (58.3\%) &  &  &  & 0.718 & 1.000 \\
NRS n=16 (50.0\%) &  &  &  &  & 1.000 \\
REV n=1 (100.0\%) &  &  &  &  &  \\
\hline
\end{tabular}
}
\subcaption{probability}
\end{subtable}
\hfill
\begin{subtable}{0.32\textwidth}
\centering
\resizebox{\linewidth}{!}{%
\begin{tabular}{lccccc}
\hline
 & \rotatebox{45}{ACC} & \rotatebox{45}{ABD} & \rotatebox{45}{CHP} & \rotatebox{45}{NRS} & \rotatebox{45}{REV} \\
\hline
CLS n=90 (77.8\%) & 0.698 & 0.059 & 0.161 & 0.115 & 1.000 \\
ACC n=60 (75.0\%) &  & 0.195 & 0.294 & 0.213 & 1.000 \\
ABD n=77 (63.6\%) &  &  & 0.755 & 0.583 & 1.000 \\
CHP n=12 (58.3\%) &  &  &  & 1.000 & 1.000 \\
NRS n=16 (56.2\%) &  &  &  &  & 1.000 \\
REV n=1 (100.0\%) &  &  &  &  &  \\
\hline
\end{tabular}
}
\subcaption{bucket\_median\_score}
\end{subtable}
\\[1em]

% --- make_and_kill_mutants_from_diff_intent_and_risks ---
\textbf{intent-aware workflow:}
\\[0.3em]
\begin{subtable}{0.32\textwidth}
\centering
\resizebox{\linewidth}{!}{%
\begin{tabular}{lccccc}
\hline
 & \rotatebox{45}{ACC} & \rotatebox{45}{ABD} & \rotatebox{45}{CHP} & \rotatebox{45}{NRS} & \rotatebox{45}{REV} \\
\hline
CLS n=46 (65.2\%) & 0.478 & 0.298 & 0.612 & 0.511 & 0.362 \\
ACC n=30 (56.7\%) &  & 0.095 & 1.000 & 1.000 & 0.452 \\
ABD n=29 (79.3\%) &  &  & 0.241 & 0.137 & 0.233 \\
CHP n=4 (50.0\%) &  &  &  & 1.000 & 1.000 \\
NRS n=11 (54.5\%) &  &  &  &  & 1.000 \\
REV n=1 (0.0\%) &  &  &  &  &  \\
\hline
\end{tabular}
}
\subcaption{rubfake}
\end{subtable}
\hfill
\begin{subtable}{0.32\textwidth}
\centering
\resizebox{\linewidth}{!}{%
\begin{tabular}{lccccc}
\hline
 & \rotatebox{45}{ACC} & \rotatebox{45}{ABD} & \rotatebox{45}{CHP} & \rotatebox{45}{NRS} & \rotatebox{45}{REV} \\
\hline
CLS n=45 (68.9\%) & $\leftarrow$\textbf{0.016} (M) & $\leftarrow$\textbf{0.005} (M) & 0.588 & 0.706 & 1.000 \\
ACC n=28 (39.3\%) &  & 0.787 & 1.000 & 0.062 & 0.414 \\
ABD n=29 (34.5\%) &  &  & 0.610 & $\uparrow$\textbf{0.025} (L) & 0.367 \\
CHP n=4 (50.0\%) &  &  &  & 0.520 & 1.000 \\
NRS n=10 (80.0\%) &  &  &  &  & 1.000 \\
REV n=1 (100.0\%) &  &  &  &  &  \\
\hline
\end{tabular}
}
\subcaption{probability}
\end{subtable}
\hfill
\begin{subtable}{0.32\textwidth}
\centering
\resizebox{\linewidth}{!}{%
\begin{tabular}{lccccc}
\hline
 & \rotatebox{45}{ACC} & \rotatebox{45}{ABD} & \rotatebox{45}{CHP} & \rotatebox{45}{NRS} & \rotatebox{45}{REV} \\
\hline
CLS n=45 (84.4\%) & 0.236 & $\leftarrow$\textbf{0.004} (M) & 0.522 & 0.365 & 1.000 \\
ACC n=28 (71.4\%) &  & 0.175 & 1.000 & 1.000 & 1.000 \\
ABD n=29 (51.7\%) &  &  & 0.607 & 0.464 & 1.000 \\
CHP n=4 (75.0\%) &  &  &  & 1.000 & 1.000 \\
NRS n=10 (70.0\%) &  &  &  &  & 1.000 \\
REV n=1 (100.0\%) &  &  &  &  &  \\
\hline
\end{tabular}
}
\subcaption{bucket\_median\_score}
\end{subtable}
\\[1em]

% --- make_tests_from_parent_and_dodgy_diff ---
\textbf{dodgy diff workflow:}
\\[0.3em]
\begin{subtable}{0.32\textwidth}
\centering
\resizebox{\linewidth}{!}{%
\begin{tabular}{lcccc}
\hline
 & \rotatebox{45}{ACC} & \rotatebox{45}{ABD} & \rotatebox{45}{CHP} & \rotatebox{45}{NRS} \\
\hline
CLS n=45 (37.8\%) & 0.163 & 1.000 & 0.065 & 1.000 \\
ACC n=32 (56.2\%) &  & 0.114 & 0.439 & 0.395 \\
ABD n=48 (37.5\%) &  &  & 0.063 & 1.000 \\
CHP n=8 (75.0\%) &  &  &  & 0.277 \\
NRS n=6 (33.3\%) &  &  &  &  \\
\hline
\end{tabular}
}
\subcaption{rubfake}
\end{subtable}
\hfill
\begin{subtable}{0.32\textwidth}
\centering
\resizebox{\linewidth}{!}{%
\begin{tabular}{lcccc}
\hline
 & \rotatebox{45}{ACC} & \rotatebox{45}{ABD} & \rotatebox{45}{CHP} & \rotatebox{45}{NRS} \\
\hline
CLS n=45 (55.6\%) & 0.817 & 0.216 & 1.000 & $\leftarrow$\textbf{0.023} (L) \\
ACC n=32 (59.4\%) &  & 0.171 & 1.000 & $\leftarrow$\textbf{0.020} (L) \\
ABD n=48 (41.7\%) &  &  & 0.445 & 0.074 \\
CHP n=8 (62.5\%) &  &  &  & $\leftarrow$\textbf{0.031} (L) \\
NRS n=6 (0.0\%) &  &  &  &  \\
\hline
\end{tabular}
}
\subcaption{probability}
\end{subtable}
\hfill
\begin{subtable}{0.32\textwidth}
\centering
\resizebox{\linewidth}{!}{%
\begin{tabular}{lcccc}
\hline
 & \rotatebox{45}{ACC} & \rotatebox{45}{ABD} & \rotatebox{45}{CHP} & \rotatebox{45}{NRS} \\
\hline
CLS n=45 (71.1\%) & 0.601 & 1.000 & 0.252 & 0.087 \\
ACC n=32 (78.1\%) &  & 0.606 & 0.182 & $\leftarrow$\textbf{0.047} (L) \\
ABD n=48 (70.8\%) &  &  & 0.254 & 0.087 \\
CHP n=8 (50.0\%) &  &  &  & 0.627 \\
NRS n=6 (33.3\%) &  &  &  &  \\
\hline
\end{tabular}
}
\subcaption{bucket\_median\_score}
\end{subtable}

\caption{P-value matrices for false positive rates (score $< 0$) comparing diff status pairs (upper triangle), grouped by catch generation approach. P-values computed using Fisher's Exact Test; effect sizes using Cohen's $h$ \cite{cohen:statistical-power}. Bold values are significant at $\alpha=0.05$. $\leftarrow P$ = row has higher FP rate; $\uparrow P$ = column has higher FP rate. Effect size (Cohen's $h$): N=negligible ($|h|<0.2$), S=small ($0.2 \leq |h|<0.5$), M=medium ($0.5 \leq |h|<0.8$), L=large ($|h| \geq 0.8$). Status abbreviations: CLS=CLOSED, ACC=ACCEPTED, ABD=ABANDONED, CHP=CHANGES\_PLANNED, NRS=NEEDS\_REVISION, REV=REVERTED.}
\label{tab:fp-rate-pvalue-matrices-by-approach}
\end{table*}

%%% END OF AUTO GENERATED TEXT ... DO NOT MANUALLY EDIT ...
%%% END OF AUTO GENERATED TEXT ... DO NOT MANUALLY EDIT ...
%%% END OF AUTO GENERATED TEXT ... DO NOT MANUALLY EDIT ...
%%% END OF AUTO GENERATED TEXT ... DO NOT MANUALLY EDIT ...
%%% END OF AUTO GENERATED TEXT ... DO NOT MANUALLY EDIT ...
%%% END OF AUTO GENERATED TEXT ... DO NOT MANUALLY EDIT ...
%%% END OF AUTO GENERATED TEXT ... DO NOT MANUALLY EDIT ...
%%% END OF AUTO GENERATED TEXT ... DO NOT MANUALLY EDIT ...
%%% END OF AUTO GENERATED TEXT ... DO NOT MANUALLY EDIT ...
%%% END OF AUTO GENERATED TEXT ... DO NOT MANUALLY EDIT ...
%%% END OF AUTO GENERATED TEXT ... DO NOT MANUALLY EDIT ...

\section{Detailed statistical analysis of catch approaches on true/false positive assessors over Human-Labelled sets of diffs}
Section~\ref{sec:stats} provided an overview of the statistical analysis of the data for catch approaches and true/false positive assessors on human labeled data.
For reasons of brevity, Section~\ref{sec:stats} provided only the primary scientific conclusions.
For completeness, and to provide more details to support our scientific conclusions and to more fully support replication, this appendix gives a more in-depth description of the statistical analysis and results.

We use  human labeled diffs, which have a final human designation, either landed into the code base or accepted to be landed, (on the positive side).
These positive (`good') labels are to be contrasted with the four `bad' statuses:  
abandoned, 
reverted,  
changes planned (recognized  by the diff author as requiring changes)
and  with
Needs revisions (where revisions are required by the diff reviewer).

Although these labels do not correspond directly to whether or not a diff contains a bug, 
let alone whether a test has caught that bug, 
they do provide a useful source of diffs with human labeled acceptance criteria.

\subsection{Methodology}
In this section we explain  the diff labelling, and the influential statistical tests we used, and how we managed the impact of multiple influential statistical tests being employed.

\noindent
{\bf Implied human `labelling' of diffs: }
We call the diffs that are accepted to land (or already landed) by human engineers `Good' diffs, and those that are neither accepted nor landed, `Bad' diffs.  
The words `Good' and `Bad' are used here only as a convenient shorthand. 
We make no claims that an individual diff labelled `good' is somehow guaranteed to be better than any labelled 'bad'.
Rather, we hope that there may be some overall statistical effect (observable at sufficient scale) 
that reflects an inherent non-random bias amongst humans when deciding whether to allow a diff to land into the code base.
This is a kind of implicit human labelling, and although it is not the ground truth for ultimate questions about true and false positives, we hope it has some (statistically discoverable) relation to these questions.

\noindent
{\bf Why this diff labelling is useful: }
If there were to be no difference between diffs we  label `Good' and those we label `Bad', 
then it would call the entire modern code review process into question. 
However, code review has proved to be one of the enduring positive results for Software Engineering research and practice.
Various forms of code review have been repeatedly found successful in Empirical Software Engineering research \cite{bacchelli:expectations,sadowski:modern} and code review  have been widely studied and shown to be correlated with the defect prevention since the 1970s and the introduction  of Fagan inspections~\cite{fagan:design}.
Based on this comprehensive previous literature on modern code review, 
we expect {\em some} statistical differences to emerge between the two broad categories of `Good' and `Bad' diffs. 

We might also expect some differences between sub-categories of diff status (e.g., `accepted' vs. `needs revision'), where there is a sufficiently large sample size to observe such effects, so we also explore this.
If we find no such statistically significant differences, then we may have not collected enough data to observe these and have insufficient statistical power.
If we observe no significant differences, and we believe statistical power to be sufficient to reveal them, then we would be forced to conclude there is no difference (or a difference with negligible effect size) between the techniques for generating weak catches, and/or that we can have no confidence in the assessments produced by the assessors that determine whether a weak catch is strong (the test failure is a true positive).

\noindent
{\bf Correlation analysis: } 
It is also interesting to speculate that there may be some correlation between the true and false positive assessors.
These are based on the enumeration of different levels of confidence, with relatively few levels, and therefore it is not appropriate to expect anything other than a rank correlation based on this ordinal scale measurement~\cite{shepperd:foundations}.
We therefore report results for Spearman and Kendall rank correlation.
We use Kendall as well as Spearman, since Spearman can be adversely affected by ties, and there are  relatively few discrete rankings employed for Rubfake and  Ensemble Categorical Likelihood, meaning that there may be a larger number of ties in the data.

The rank correlation investigates the degree to which the three different techniques for assessing true/false positives would agree on an ordering. 
However, these techniques only produce orderings as an indication of their confidence. 
Ultimately, they are attempting to approximate a binary classification problem, 
since either a test failure is a true positive or it is a false positive.

We treat the classification as two separate classification problems.
One is the problem of determining whether a test failure is a true positive, 
while the other is that of determining whether it is a false positive.
We formulate the overall problem in these two distinct ways, 
because the rankings can give the answer zero (neither the true positive nor false positive). 
We  only act on any assessment when it gives a confident prediction of one or the other,
treating a zero assessment as `no information'.

We complement the correlation analysis with a report on the degree of inter-rater agreement between the classifiers, both collectively, and pairwise.
If it turns out that the classifiers are strongly correlated (and/or have high inter-rater agreement), 
then this would suggest that each is merely a proxy of the other assessors. 
However, if they exhibit low correlation, then they may be complementary. 
The ultimate decision about whether a test failure is true or a false positive 
would best be determined by some {\em aggregated} decision process that takes {\em all} of these different assessors into account.

\noindent
{\bf Analysis of statistic significant differences: }
We investigate statistical differences between the predictions of true positive, respectively false positive, between the two categories of diffs (`Good' vs, `Bad'), and also between sub categories of diff status label.
For this comparison we use Fisher's exact test and, where this produces a significant outcome at the $\alpha = 0.05$ level, we also report the effect size, using Cohen's $h$ statistic.
For the $h$ statistic, we report  the  effect size `bucket' according to the (widely-used) rule-of thumb suggested by Jacob Cohen himself~\cite{cohen:statistical-power}.
That rule-of-thumb buckets an effect size as follows: 
`Small' ($h \ge 0.2)$, 
`Medium' ($h \ge 0.5$) 
and 
`Large' ($h \ge 0.8$).
When the result is significant according to the Fisher test, 
and Cohen's $h$ is less than $0.2$, 
we deem the effect size to reside in the `Negligible' bucket.

\noindent
{\bf Catering for spurious statistical significance: }
Since we are executing multiple statistical tests, there is a probability that some $P$ values that are found to be significant at the $\alpha = 0.05$ level  will be spurious.
After all, the  $\alpha$ level $0.05$ simply means that we have at most a 1 and 20 chance of a spurious result.
However, we do not want to adopt an arbitrary  conservative $p$-value correction approach, such as Bonferroni (nor even the, less conservative, Hochberg~\cite{hochberg:sharper} correction), 
since we are simply looking for an understanding of whether  there is some form of 
distinction between  behaviors on different classes of diffs according to human labeling.
We are not using the outcomes to make life-altering decisions about drug treatments~\cite{gelman2012why,perneger1998whats,rothman1990no}.

As with most scientific work, and certainly almost all Software Engineering applications~\cite{mhetal:sbse-tutorial,arcuri:practical} it is not just statistical significance that matters, but the effect size; since we can increase the sample size simply by deploying additional computational resources, 
we are likely to be able to find significant results, even where the effect size is negligible.

We are thus, obviously, more interested in those results where the effect size is larger.
With this in mind, we assess the likelihood of obtaining a statistically significant result 
with the given effect size or higher, using similar data to that we study.
We achieve this comparison using sample sizes that are approximately the same size as those we study, but simply discarding labels and thereby choosing arbitrary sets of data for each of the two simulated categories.
With these arbitrary sets of data, we can reliably estimate the chance of obtaining the statistically significant result at or above a given effect size for the data used in this paper.
Our approach is essentially an implementation of a permutation-based resampling procedure~\cite{good2013permutation}.

\subsection{Results}
Overall, based on the results of the analysis, 
we believe that the `Dodgy diff' workflow does, indeed, generate more tests, 
and likely also it generates more false positives.
Although it generates more tests overall, this does not mean it is able to catch bugs in more diffs, 
nor even that it generates tests for more diffs.
Indeed, as shown in Table~\ref{tab:numbers-of-weak-catch-results} the `Dodgy diff' workflow tends to generate many tests for a single diff, rather than covering more diffs overall, when compared to the intent-aware workflow.

We also believe that there is reason to have confidence that all three of the true/false assessors have {\em some} distinguishing power between `Good' and `Bad' diffs, 
both when predicting true positive and when predicting false positive.
That is, the differences observed between `good' and `bad' diffs suggest that the assessments are, at least, 
generally consistent with the human diff labeling.
Taken together with the results for correlation, which indicate only weak correlation between them, we believe that this gives us overall confidence in our assessors, but suggests that they each contribute a different facet of the overall decision. 
Therefore, they need to be used in concert to arrive at a final decision about the likelihood of true/false positive test failure.

\subsubsection{Permutation-based resampling}
Table~\ref{tab:fisher-randomization-sanity-check} presents the results of the Fisher's Exact Test randomization sense check, where 100,000 tests were performed on randomly shuffled data (effectively discarding diff labels to make arbitrary simulated `Good' and `Bad' diff sets). 
With $\alpha = 0.05$, we expect approximately 5,000 spuriously significant outcomes per metric by chance ($alpha = 0.05$).
This table is used to assess the likelihood of achieving a given minimal effect size for the test we perform on diffs
labelled `Good` and `Bad`.
As can be seen from the table, the chance of obtaining a significant result with small effect size is approximately similar to the $alpha$ level traditionally used for significance (only slightly lower than 0.05).

With more than 20 statistical tests, we can therefore expect to find some spurious significant results with small effect size.
By contrast, the chance of obtaining a significant result with medium or large size is close to  zero.
Therefore, we should treat small effect size results with a great degree of caution, 
but can have considerably more confidence in those with medium or large size, 
based on this size of sample between `Good' and `Bad'.

These results are in stark contrast to those for smaller sample sizes.
That is, Table~\ref{tab:fisher-randomization-small-sample}  was motivated by 
the fact that the sample sizes were so much smaller than those covered by the permutation-based sampling in Table~\ref{tab:fisher-randomization-sanity-check}, thereby elevating the chance of a spurious statistical inference.  

With $\alpha = 0.05$, we expect approximately 5,000 false positives per metric by chance ($alpha = 0.05$), as before.
Indeed, with such small sample sizes, the chances of a significant result are somewhat diminished, as the table shows.
Nevertheless, and much more concerning, notice the relatively high  (compared to Table~\ref{tab:fisher-randomization-sanity-check}) probability of obtaining results that are inferred to exhibit either a medium or a large effect size.

This suggests that  a {\em great deal of caution} is required when interpreting any results for subclasses of diff status. 
Some of these are inherently rare, such as diffs which are reverted, 
and therefore, even at Meta scale, 
we may not have sufficiently many data points for the statistical power required to make confident claims.
Nevertheless, for completeness and replicability, in this appendix, we present all of these results.

\subsubsection{Correlation and Inter-rater Agreement Between True/false Positive Assessors}
Table~\ref{tab:agreement-and-correlation} presents the inter-rater agreement between the assessors and the results of the rank correlation and analysis.
In both tables, we use a background gray scale to indicate the strength of correlation/agreement, 
with darker shades of gray indicating stronger correlation/agreement.
As can be seen, there is a greater degree of correlation and greater agreement between the two LLM-based measures than there is between either and the rule-based approach.
This is a reflection of the fact that the rule-based approach uses an inherently fundamentally 
different style, and suggests that it is highly complementary to any LLM-based approach.

It is interesting that the highest degrees of inter-rater agreement, 
between all the assessors together, 
and between each pairwise, 
are observed for the category of diffs marked as `Needs Revisions (NRS)'. 
This class of diffs is that for which we have the greatest confidence that the human labeling signifies the presence of an issue,  because a human reviewer has specifically identified that the diff in question needs further revision.

\subsubsection{Significant differences between diff status categories for workflows and true/false positive assessors}
Table~\ref{tab:approach-pvalues} gives the top level results for the statistical differences between diffs labelled `Good' and `Bad' for each of the two workflows and each of the three true/false positive assessors, 
based on whether they predict a true positive or a false positive for each test failure.
As can be seen, the `dodgy diff' workflow tends to lead to test failures for which all assessors produce a more noticeable difference, both when predicting true positive and when predicting false positive.

Given the results of the  permutation based sampling in Table~\ref{tab:fisher-randomization-sanity-check} 
the joint probability that these observations could have resulted merely by chance is extremely low.
The `Dodgy diff' workflow makes no attempt to produce a true positive, 
but merely seeks to maximize the generation of tests which simply {\em distinguish} between the observable behaviors of the 
parent of the diff and the tested diff.

This makes it more likely that it will produce false positives, and it is reassuring that the three assessors appeared to behave significantly differently when applied to diffs labelled `Good' and `Bad' in this regard.
It also meets with the underlying  intuition that the assessors tend to be more likely to assess a test failure as a true positive on diffs labelled `Bad', and more likely to assess it to be a false positive on the diffs labelled 'good'.
Although not entirely sufficient on its own, this is the strongest statistical evidence we have that the assessors give a reasonable signal,  that is significantly better than random guessing.

Furthermore, they also behave significantly differently with regard to true positives, and for all three assessors.
This is important because good performance on predicting false positives will help us to remove strictly weak catches and avoid unnecessary drag on developer velocity.
Essentially, the importance of this is to improve the precision of the overall catching technique.
However, if we can find techniques that also have high probability of predicting {\em true} positive, then we will also improve the recall of the overall catching approach; catching more bugs just in time.

Tables~\ref{tab:tp-rate-pvalue-matrices-by-approach} and \ref{tab:fp-rate-pvalue-matrices-by-approach} show the statistical significance and effect size for differences between each pairwise diff status comparison.
With this many inferential statistical tests, we should be very careful to guard against spuriously significant outcomes.
In particular, even though we noticed some effect sizes which are `Large' (according to traditional interpretations of  Cohen's $h$ statistic), 
we need to recall that there is a non-trivial probability of this occurring by chance according to the permutation-based resampling results from Table~\ref{tab:fisher-randomization-small-sample}. % I dont understand why Overleaf renders this as ??
We include these results primarily for completeness and replicability, 
and therefore refrain from drawing any scientific conclusions at this stage.

\subsubsection{Significant differences between weak catch generator rates}
Table~\ref{tab:catch-rate-approaches} uses the Fisher exact test and Cohen $h$ to investigate whether there are significant differences with non-trivial  effects sizes  between the catch rates of the two techniques.
Overall, there is little evidence for any significant differences.
This could be an instance of Simpsons paradox \cite{simpson1951interpretation,pearl2009causality}, but given the results from the permutation study, it could also result from a lack of sufficient statistical power given these relatively small sample sizes for individual diff statuses.

Where there is a difference, it tends to suggest that each technique has a slightly higher catch rate for closed diffs (that have been accepted and landed into production), than for the other diff status classes.
Since the effect size is negligible or small, this is only a very slight effect, into which we should not read too much.
If indeed there is any difference, the likely explanation is that these closed diffs denote completed changes, 
whereas abandoned diffs and those for which changes are planned, 
for example, may only be partially written, therefore with fewer changes at the time the test was generated and therefore fewer opportunities to generate distinguishing tests between parent and child.

\end{document}